\documentclass[%
notitlepage,
nofootinbib,
amsmath,
amssymb,
aps,
twocolumn
]{revtex4-1}

\usepackage{graphicx}
\usepackage{dcolumn}
\usepackage{bm}
\usepackage{mathtools}
\usepackage{amsmath}
\usepackage{amssymb}
\usepackage[hidelinks]{hyperref}
\usepackage{setspace}

\newcommand{\rb}{\bar{r}}

\begin{document}

\title{The rotating black hole interior: Insights from \\
	gravitational collapse in $AdS_3$ spacetime}

\author{Alex Pandya}
\email{apandya@princeton.edu}
\author{Frans Pretorius}
\email{fpretori@princeton.edu}
 \affiliation{Department of Physics, Princeton University, Princeton, New Jersey 08544, USA.}

\date{\today}

\begin{abstract}
We present results from a numerical study of rotating black hole formation in 3-dimensional asymptotically anti-de Sitter (AdS) spacetime, focusing on the structure of the black hole interior. While black holes in $AdS_3$ are of theoretical interest for a wide variety of reasons, we choose to study this system primarily as a toy model for astrophysical (4-dimensional) black holes formed from gravitational collapse. We investigate the effect of angular momentum on the geometry inside the event horizon, and see qualitative changes in the interior structure as a function of the spin parameter.  For low spins, we find that a central spacelike curvature singularity forms, connecting to a singular, null Cauchy horizon. For spins above a threshold consistent with the linear analysis of Dias, Reall and Santos, curvature on the Cauchy horizon remains bounded, signaling a violation of the strong cosmic censorship conjecture. Further increasing the spin leads to a decrease in the relative size of the spacelike branch of the singularity, which vanishes completely above a second threshold. In these high-spin cases, the interior evolution is bounded by a regular Cauchy horizon, which extends all the way inward to a regular, timelike origin. We further explore the geodesic focusing (``gravitational shock-wave'') effect predicted to occur along the outgoing branch of the inner horizon, first described by Marolf and Ori.  Remarkably, we observe the effect at late times in all of the black holes we form, even those in which the inner apparent horizon collapses to zero radius early in their evolution.
\end{abstract}

\maketitle

\section{Introduction} \label{sec:introduction}
If governed by general relativity, the exterior structure of a sufficiently isolated black hole is expected to be well described by a member of the celebrated Kerr family of solutions~\cite{kerr63}. The historic LIGO measurement of gravitational waves from the merger of two black holes~\cite{abbott2016} has given the first quantitative evidence supporting this expectation, which is also consistent with the first image of a black hole taken by the EHT Collaboration~\cite{akiyama2019}.  

Given the aforementioned results, it is natural to ask why the Kerr model has such relevance for astrophysics, especially since it possesses a high degree of symmetry (axisymmetry, stationarity) not present in any natural setting. Several properties of strong-field general relativity in 4-dimensional (4D) spacetimes give the answer: (1) the only stationary black hole solutions in vacuum are the Kerr family (the ``no-hair theorems''~\cite{israel,1971PhRvL..26..331C,hawking-uniqueness,1975PhRvL..34..905R}); (2) when gravitational collapse occurs, the singularities that necessarily form are hidden behind an event horizon (Penrose's weak cosmic censorship conjecture); and (3) dynamical perturbations of the black hole exterior geometry always decay, and either fall into the black hole or are radiated away. The third effect is especially strong, since the ``perturbations'' can initially be arbitrarily large (case in point the collision of two black holes), transitioning to the linear regime of exponential quasinormal mode decay, followed at late times by a power law decay~\cite{price1972}. Taken all together, these properties are sometimes referred to as the {\em final state conjecture}~\cite{penrose1982}, or the conjectured nonlinear stability of the Kerr solution, a mathematical proof of which remains elusive.

The features that allow the Kerr geometry to be so successful in modeling the black hole exterior do not have the same effect on the interior. Crucially, there is no ``decay'' property which will allow the interior to asymptotically approach a unique end state; as a result, the detailed structure inside any particular black hole will depend strongly on the properties of the matter that collapsed to form it.  This dependence on the black hole's history endows the interior with a much richer structure, though at the cost of increased mathematical complexity, and a number of interesting problems remain unsolved.

Perhaps one of the most pressing open questions about the black hole interior is the generic nature of the singularity (or whatever form of spacetime incompleteness we know must be present~\cite{Penrose:1964wq}). Early on in the study of interiors, the similarity of the Schwarzschild interior to a Kasner spacetime, together with the relevance of the latter in the analysis of Belinksi, Lifschitz and Khalatnikov (BKL)~\cite{1971ZhETF..60.1969B} on singularities in so-called cosmological spacetimes, spurred many to argue such a spacelike curvature singularity would denote the classical end of the generic black hole interior~\cite{2002LRR.....5....1B,Garfinkle:2003bb}. 

\begin{figure*}[]
	\centering
	\includegraphics[width=\textwidth]{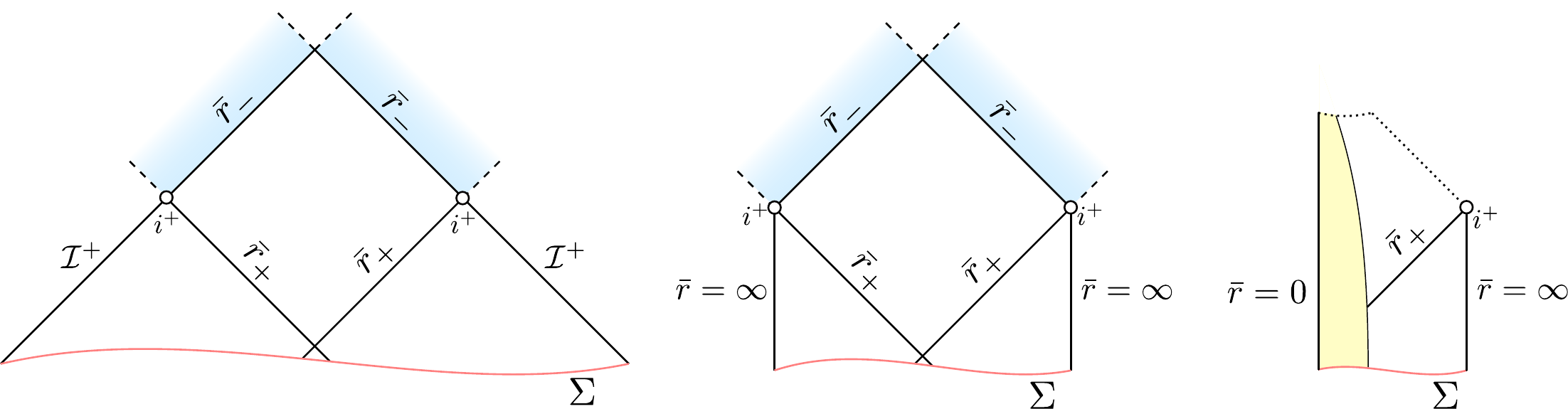}
	\caption{Spacetime (Penrose) diagrams illustrating the Kerr geometry (left panel), BTZ geometry (center panel) and a possible dynamical rotating collapse geometry in asymptotically AdS spacetime (right panel). In the 4D Kerr case, each point on the diagram represents a 2-sphere geometry with proper area $\propto \rb^2$, and for the 3D cases each point represents a circle with proper circumference $\propto \rb$. In each of the two eternal black hole cases the spacetimes beyond $\rb = \rb_-$ contain causal curves that terminate at the dashed lines, never intersecting the red Cauchy surface $\Sigma$.  This fact implies that the blue (gradient) shaded regions cannot be uniquely evolved from the earlier spacetimes without somehow specifying data on the dashed lines, and thus $\rb = \rb_-$ is a Cauchy horizon.  Note also the similarities between the Kerr and BTZ spacetimes, which differ only at $\rb = \infty$.  Future timelike infinity $i^+$ is marked on each diagram; on the right panel the solid yellow region represents collapsing matter, and spacelike/null singularities to its causal future are denoted by dotted lines.}
	\label{fig:penrose_diagrams}
\end{figure*}

The reliance of these arguments on the spherically symmetric Schwarzschild solution, however, significantly weakens their claims to the structure inside realistic black holes.  Relaxing the symmetry to axisymmetry, one arrives at the aforementioned Kerr geometry, which is vastly different from Schwarzschild in the interior for any nonzero dimensionless spin parameter $a$. In particular, in Kerr there is a null Cauchy horizon that is expected to become singular when subject to the perturbations present in a realistic collapse.  Some have argued such a null singularity should be part of the generic black hole interior\footnote{Note that the ``textbook'' timelike singularity in Kerr is more an artifact of demanding an analytic extension of the metric across the Cauchy horizon, and is not expected to be relevant in any collapse solution of the interior.}\cite{Ori:1995nj,Luk:2013cqa}; others have countered that due to the nonlinearity of the field equations, a spacelike singularity might form in the interior well before any Cauchy horizon, restoring the BKL picture even in rotating collapse. 

A recent breakthrough by Dafermos and Luk, however, has proven otherwise~\cite{dafermos2017}, showing that if the exterior stability of Kerr is assumed, the piece of the Cauchy horizon ``connecting'' to it (on a Penrose diagram---see the left panel of Fig.~\ref{fig:penrose_diagrams}) will always be present. Moreover, this branch of the Cauchy horizon will be ``weakly singular'' in the sense that although a curvature singularity is present, the metric itself is well defined there and can be extended continuously across it. Thus, the $C^0$-inextendible formulation of Penrose's strong cosmic censorship conjecture~\cite{penrose1978} is false in this case, though a weaker version such as that of Christodoulou's~\cite{christodoulou2008} likely holds (for a comprehensive discussion of the cosmic censorship conjecture in this context see the introduction of~\cite{dafermos2017}, and references cited therein).

One problem hampering the development of a more complete understanding of the realistic black hole interior is that there are no explicit solutions known for guidance, numerical or otherwise. For numerical evolution, one difficulty in obtaining such solutions is the lack of symmetries that can be applied in the generic case, and this, together with the rather extreme spacetime dynamics expected to unfold, makes it unclear what coordinate conditions to impose in order to reveal the full Cauchy development of relevant initial data. For example, any observer reaching the Cauchy horizon will do so in finite proper time, at which point the {\em entire} future evolution of the full exterior spacetime must be complete, as by then it will be in the past domain of dependence of these observers. To circumvent such difficulties, earlier studies have either focused on spherically symmetric charged collapse as a toy model for Kerr~\cite{brady1995,hod1998,hod1999,chesler2019}, or ignored the collapse and perturbed about a segment of the inner horizon of Kerr~\cite{chesler2018}. 

The reason charged collapse is used as a toy model for rotating collapse is that the analogous Reissner-Nordstrom black hole solution has a similar Penrose diagram to Kerr, including a null Cauchy horizon (the left panel of Fig.~\ref{fig:penrose_diagrams}), but in contrast to Kerr, formation of a charged black hole can be studied in spherical symmetry. This additional symmetry offers many simplifications for both numerical and analytical studies. For numerics, spherical symmetry makes it straightforward to construct global coordinates adapted to the radial causal structure of the spacetime, and that map all the relevant infinities on a Penrose diagram to finite grid locations. What numerical studies of charged scalar field collapse in spherical symmetry have revealed in a handful of select cases~\cite{brady1995,hod1998} is that a curvature singularity forms in the interior which has a central spacelike branch connected to a singular Cauchy horizon, giving a Penrose diagram similar to that on the right panel of Fig.~\ref{fig:penrose_diagrams} (except these studies have been in asymptotically flat 4D spacetime, replacing the timelike infinity of AdS with null infinity). The singularity on the Cauchy horizon is ``mild'' in the sense of tidal forces there~\cite{Ori:1992zz}, and exhibits the ``mass inflation'' phenomenon first discovered by Poisson and Israel~\cite{poisson1990} (see also~\cite{Dafermos:2003wr}) where the Hawking quasilocal mass function diverges.

Another interesting question about the interior in a collapse scenario involves what happens to the inner horizon (the left branch of $\bar r_-$ on the left panel of Fig.~\ref{fig:penrose_diagrams}): does it form, and are similar pathologies present there as in the approach to the Cauchy horizon? One could expect problems, as the inner horizon is the Cauchy horizon from the perspective of the ``other universe'' in an eternal black hole spacetime. On the other hand, what governs the stability and regularity of the Cauchy horizon is ultimately the influx of radiation from the exterior, which is controlled by the essentially unique decay rates outside the black hole. Studies of perturbations of the inner horizon have not been subject to such strong guidance on appropriate initial data, and more {\em ad hoc} prescriptions have been used. Marolf and Ori~\cite{marolf2012} first explored this region of the interior at the perturbative level, finding that an observer crossing the inner horizon would be met with an extremely rapid variation in the  metric, stress-energy, and curvature near it, likely being destroyed by diverging tidal forces \textit{before} meeting the singularity. They dubbed this a null shockwave singularity, though showed it only  becomes a true ``shock'' in the sense of a discontinuity in the metric when it reaches the Cauchy horizon. Numerical studies of similar setups about the Reissner-Nordstrom~\cite{eilon2016,Eilon:2016bcl,chesler2019,Chesler:2020lme} and Kerr inner horizons~\cite{chesler2018} confirmed this result at the nonlinear level, though the evolutions could not proceed all the way to the Cauchy horizon. 

\begin{figure*}[]
	\centering
	\includegraphics[width=\textwidth]{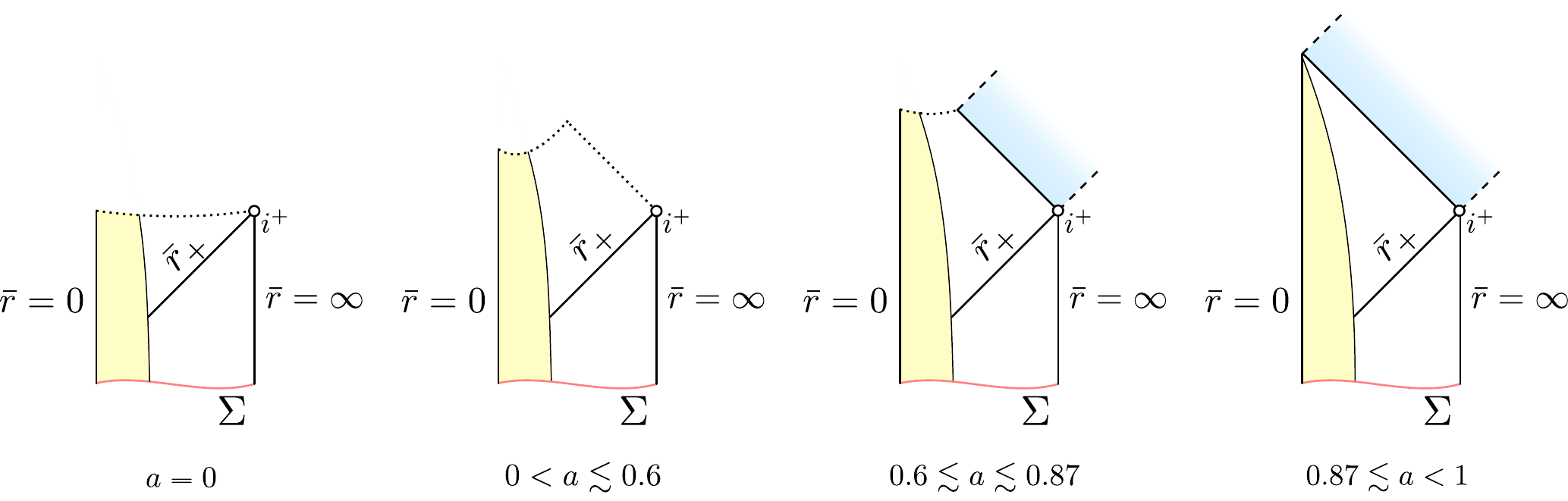}
	\caption{Penrose diagrams for the $AdS_3$ black hole spacetimes we observe in our numerical solutions (similar annotation as explained in the caption of Fig.~\ref{fig:penrose_diagrams}, and in contrast to the eternal BTZ black hole shown in the middle panel there).  Note that the structure interior to the event horizon is strongly dependent on the value of the spin parameter $a$.  See Sec.~\ref{sec:bh_interior} for a detailed explanation.}
	\label{fig:a_dep_penrose_diagrams}
\end{figure*}

In this work we study a different toy model for the realistic black hole interior: formation of rotating black holes in 3D asymptotically $AdS$ spacetime ($AdS_3$). This model shares two of the main features Reissner-Nordstrom offers as a (potentially) useful analogue of Kerr: the Penrose diagrams are similar (Fig.~\ref{fig:penrose_diagrams}), and the problem can be explored in circular symmetry (the analogue in 3D of spherical symmetry in 4D). An advantage over the charged collapse models is that here we use rotating matter, and can study how angular momentum itself affects the interior structure\footnote{An added benefit of this model is its potential relevance to string theory  and conformal field theories (CFTs) through the AdS/CFT correspondence (see e.g. \cite{skenderis1999,birmingham2001,Witten:2007kt,delafuente2014}), though we do not explore that here.}. However, there are several differences in 3 vs 4-dimensional gravity that bare keeping in mind in anticipating how closely the 3D model might be able to capture qualitative features of the 4D case. Key among these are that in 3D a negative cosmological constant is required for black hole solutions~\cite{ida2000}, that 3D Einstein gravity does not admit a Newtonian limit~\cite{1986CQGra...3..551B}, and furthermore that there are no freely propagating (gravitational wave) degrees of freedom, as the Weyl tensor is identically zero and matter fully constrains the dynamics and curvature through the Einstein equations.

It is therefore quite surprising that an analogue to the Kerr solution even exists in 3D, namely the celebrated Ba\~nados, Teitelboim, and Zanelli (BTZ) black hole solution \cite{Banados:1992wn}, which moreover shares many of the properties of Kerr black holes~\cite{banados1993,carlip1995}. In a rough sense one can envision the geometry of a BTZ black hole as close to that of an equatorial slice of the Kerr geometry, which is why adding rotation to a nonrotating BTZ black hole does not break circular symmetry, and allows the problem to be studied with a (1+1)-dimensional numerical code. 

In order to dynamically form BTZ-type black holes, we choose to use a scalar field as the matter source. However, for a scalar field to carry angular momentum it cannot be circularly symmetric; we bypass this difficulty with a common ``trick'' by using a complex scalar field, and arranging for the real and imaginary components to individually have azimuthal dependence, but be out of phase to give a net circularly symmetric (real) stress-energy tensor. Regarding earlier work within this gravity plus matter model, critical collapse and the interior of nonrotating black holes was studied in~\cite{pretorius2000} (see also~\cite{Husain:2000vm}), and critical collapse of rotating black hole formation in~\cite{jalmuzna2015}. Our formalism and code are based on the latter, with extensions to be able to explore the black hole interior.

Fig.~\ref{fig:a_dep_penrose_diagrams} gives a pictorial summary of our main results. For low spins, much like in the 4D charged collapse case, we find that a curvature singularity forms that is composed of a central spacelike branch connected to a null branch. However, as the spin increases, the ``strength'' of the singular behavior on the null branch decreases, and above a dimensionless spin of $a\sim 0.6$ the Cauchy horizon ceases being singular. This is consistent with a recent linear analysis of perturbations of the Cauchy horizon of the BTZ black hole carried out by Dias, Reall and Santos~\cite{dias2019}. We also find that as the spin increases, the size of the spacelike branch on the Penrose diagram shrinks, eventually vanishing for spins above $a\sim 0.87$. Thus for such rapidly spinning black holes the interior evolution ends along a regular Cauchy horizon that extends all the way to a regular, timelike origin. In all cases, the exterior appears to asymptote to a stationary BTZ solution. In the interior an inner horizon does form, though it is never stationary, and for slowly rotating black holes it moves inward and terminates at the spacelike singularity. Nevertheless, in all cases we see an outgoing shocklike feature form in the interior as found by Marolf and Ori. 

The remainder of this work will begin with a brief overview of the BTZ spacetime and the Marolf-Ori focusing effect (Sec. \ref{sec:AdS3_and_BTZ}), the system of equations we solve (Sec. \ref{sec:mEKG}), and then our numerical algorithm (Sec. \ref{sec:numerical_method}).  This section is followed by our results (Sec. \ref{sec:results}), and we conclude with a discussion (Sec. \ref{sec:conclusion}). We leave the full expressions of the equations we solve, and some convergence results, to the appendix (Secs. \ref{sec:EOM}-\ref{sec:convergence_tests}). We use geometric units where the speed of light $c=1$ and Newton's constant $G=1/2$, and use the $-++$ signature for the metric tensor. Unless otherwise stated, we will use a prime ($'$) to denote the ordinary partial derivative of a function $f(t,r)$ with respect to the radial coordinate $r$, and similarly the overdot ($\dot{\ }$) for the partial with respect to coordinate time $t$, i.e. $f'\equiv\partial f(t,r)/\partial r$ and $\dot{f}\equiv\partial f(t,r)/\partial t$.

\section{The BTZ Geometry} \label{sec:AdS3_and_BTZ}

The Einstein field equations with a negative cosmological constant $\Lambda\equiv-1/\ell^2$ are
\begin{equation}\label{efe}
R_{\mu\nu}-\frac{1}{2} g_{\mu\nu} R + \Lambda g_{\mu\nu} = \kappa T_{\mu\nu},
\end{equation}
where $R_{\mu\nu}$ is the Ricci tensor, $R$ the Ricci scalar, $g_{\mu\nu}$ the metric tensor, $T_{\mu\nu}$ the stress-energy tensor of matter, and $\kappa$ a coupling constant (that we set to $4\pi$). 
The BTZ black hole is a solution to the above with $T_{\mu\nu}=0$. Using the analogue of Boyer-Lindquist coordinates, the line element of the BTZ geometry can be written as
\begin{equation}\label{eq:BTZ}
ds^2 = - \bar{f} dt^2 + \bar{f}^{-1} d\rb^2 + \rb^2 (d\theta + \bar{\beta} dt)^2,
\end{equation}
with
\begin{equation}\label{eq:BTZ_quantities}
\begin{aligned}
\bar{f} &= -M + \frac{\rb^2}{\ell^2} + \frac{J^2}{4 \rb^2} \\
\bar{\beta} &= -\frac{J}{2 \rb^2}.
\end{aligned}
\end{equation}
Here $M$ and $J$ are the black hole mass and angular momentum, respectively (and note that in 3D gravity $M$ is dimensionless, while $J$ has dimension of length). This line element represents a space of constant curvature (e.g. the Kretschmann scalar $K \equiv R^{\mu \nu \rho  \sigma} R_{\mu \nu \rho \sigma}$ evaluates to $K=12/\ell^4$), and all of the nontrivial causal structure encoded in the metric can be considered topological in nature\footnote{In fact, one way to derive the BTZ solution is by making appropriate identifications within AdS spacetime~\cite{banados1993}, and even more complicated multi-black hole/wormhole solutions can be constructed in this manner~\cite{Aminneborg:1997pz}.}. 
Without spin, the metric above only describes a black hole spacetime for $M > 0$; the $M = -1$ case is pure $AdS_3$, and $-1 < M \leq 0$ correspond to spacetimes with naked conical singularities at the origin. Hence if we want to form a black hole by gravitational collapse beginning from regular initial data (in particular, data with no initial conical singularity), a finite amount of total mass in matter energy is required to lift the asymptotic spacetime mass above $M=0$.

Including spin $a \equiv \frac{|J|}{M \ell}$, the BTZ solution admits a number of features analogous to those of Kerr.  Among these is that there is an extremal limit $a=1$ above which there are no horizons; below this limit, the two horizons are distinct and are located at
\begin{equation} \label{eq:rbar_pm}
\rb_{\pm} = \sqrt{\frac{M \ell^2}{2} \Big( 1 \pm \sqrt{1 - a^2} \Big)}.
\end{equation}
Also as with Kerr, BTZ possesses an ergoregion between the event horizon at $\rb_+$ and the circle with radius $\rb_{erg}=\ell \sqrt{M}$; within this region all causal curves are required to rotate about the black hole in the same sense as its spin. BTZ black holes also admit a no-hair theorem~\cite{birmingham2001}---which is important for the structure of the Cauchy horizon (see Sec. \ref{sec:introduction})---among a number of other interesting results; see~\cite{banados1993} and the reviews in~\cite{birmingham2001,carlip1995,mann1995}. 

To close this section, we will outline a derivation of the Marolf-Ori focusing effect along the inner horizon of an eternal BTZ spacetime, as this analytic result will be useful to compare with our subsequent numerical results. The question here is the following: suppose there is a pulse of outgoing radiation (always coming from matter in the 3D case) propagating near the inner horizon; how is this pulse perceived by an infalling observer at late times? Specifically, how will the observed profile of the pulse change as a function of when the infalling observer crosses the event horizon, as measured by an external timekeeper? 

For simplicity, we will begin the calculation by considering our observers to be ingoing null geodesics with zero angular momentum (in the code we study both timelike and null observers). To that end, we rewrite the BTZ metric in ingoing Eddington-Finkelstein coordinates, which are regular at both horizons, by transforming to a null coordinate $v$ and new angular coordinate $\tilde{\theta}$:
\begin{equation}
dt=dv - \frac{d\rb}{\bar{f}}, \ \ \
d\theta=d\tilde{\theta}+\frac{\bar{\beta}}{\bar{f}} d\rb
\end{equation}
giving
\begin{equation} 
ds^2=-\bar{f} dv^2 + 2 dv d\rb + \rb^2 \left( d\tilde{\theta} + \bar{\beta} dv\right )^2.
\end{equation}
For the remainder of this section let an overdot denote the derivative with respect to affine parameter $\lambda$, i.e. $\dot{()}\equiv d(\ )/d\lambda$. A zero angular momentum geodesic has $\dot{\tilde{\theta}}=-\bar{\beta}\dot{v}$; of these, ingoing null geodesics are $v={\rm const.}$ curves (with $\dot{\rb}={\rm const.}$), and outgoing null geodesics satisfy
\begin{equation}
2\ddot{v} + \frac{\partial \bar{f}}{\partial \rb} \dot{v}^2 = 0, \label{geod_1}
\end{equation}
of which a first integral can be written as 
\begin{equation}
\frac{dv}{d\rb}=\frac{2}{\bar{f}}. \label{geod_2}
\end{equation}
Our numerical code does not use ingoing Eddington-Finkelstein coordinates, and of course the metric will not be the exact BTZ spacetime, so some care must be taken in defining quantities that can be meaningfully compared. To do so, we integrate affine time $\lambda_+$ along the outgoing null generator of the event horizon $\rb=\rb_+$; this is unique up to an overall constant scale and shift. We will then define the time parameter $\tilde{v}$ to be the affine time $\lambda_+$ at which the infalling geodesic crosses the event horizon. For the BTZ spacetime above, it is straightforward to find the relationship between affine parameter and Eddington-Finkelstein coordinate $v$ along either horizon from (\ref{geod_1}):
\begin{equation}
\lambda_\pm \propto e^{\pm \kappa_\pm v}, \label{affine_rpm}
\end{equation}
where $\kappa_\pm= (\rb_+^2 - \rb_-^2)/(\ell^2 \rb_\pm)$ is the surface gravity on the corresponding horizon (note also that from (\ref{affine_rpm}) it is clear that inner horizon generators reach the Cauchy horizon as $v\rightarrow\infty$ in finite affine time $\lambda_-$). Thus we define 
\begin{equation}
\tilde{v} = N (e^{\kappa_+ v}-1),
\end{equation} 
where $N$ is some (arbitrary) overall constant scale, and we (arbitrarily) set $\tilde{v}(v=0)=0$.

At an initial time $v=0$, let the extent of our outgoing test pulse range in proper circumference from $\rb_-$ to $\rb_- + \delta \rb_0$ (to which side of $\rb_-$ the pulse is, i.e. the sign of $\delta\rb_0$, does not matter). Then, from (\ref{geod_2}), it is possible to compute the extent of the pulse $\rb_- + \delta \rb$ at some later time $v$; to leading order in $\delta \rb_0/\rb_-$ it is
\begin{equation} \label{eq:delta_rb_v}
\delta\rb \sim \delta \rb_0 e^{-\kappa_- v}.
\end{equation}
This result shows the sharpening of the pulse is exponential in ingoing Eddington-Finkelstein time, at a rate controlled by the surface gravity of the inner horizon. In terms of the time $\tilde{v}$ (defined above) that we compute in the code, (\ref{eq:delta_rb_v}) translates to the following power law relationship:
\begin{equation}\label{null_steep}
\delta\rb \sim \delta \rb_0 (\tilde{v}/N+1)^{-\kappa_-/\kappa_+}.
\end{equation}

The steepening of features implied by (\ref{null_steep}) is the analogue of the blueshift effect on the Cauchy horizon, and one can anticipate that it could have a similar drastic backreaction on the geometry, provided an inner horizon of similar structure forms during collapse. Marolf and Ori considered this possibility and suggested that when backreaction is taken into account, features of the geometry are similarly focused, effectively producing an asymptotically divergent tidal force experienced by observers crossing $\rb_-$. Thus even though they argued that this ``gravitational shock-wave'' never becomes a true curvature singularity until it reaches the Cauchy horizon, at late times it is nevertheless just as disastrous to an infalling observer (or perhaps even more so, depending on how the spacetime extends across the Cauchy horizon).

\section{The Einstein-Klein-Gordon System in Asymptotically $AdS_3$ Spacetime} \label{sec:mEKG}

As described in Sec. \ref{sec:introduction}, it is possible to solve for the formation of rotating black holes in $AdS_3$ while retaining the circular symmetry of the governing partial differential equations (PDEs).  In order to do so, we follow the work of Jalmuzna and Gundlach \cite{jalmuzna2017}, and study the dynamics of a spacetime with the following metric ansatz
\begin{equation} \label{eq:jg_metric}
ds^2 = f (-dt^2 + dr^2) + \rb^2 (d\theta + \beta dt)^2,
\end{equation}
where $f \equiv e^{2 A(t,r)}/\cos^2(r/\ell)$, proper circumference $\rb \equiv e^{B(t,r)} \ell \tan(r/\ell)$, and $\beta(r,t)$ is also in general a function of $r$ and $t$. Pure AdS spacetime is given by the limit $A, B, \beta \to 0$. The radial coordinate $r \in [0, \frac{\ell \pi}{2}]$ is compactified, with timelike infinity reached in the limit $r\rightarrow \ell \pi/2$. That the $(r,t)$ sector of the metric is conformal to Minkowski spacetime $-dt^2 + dr^2$ then also implies the timelike coordinate $t$ is similarly compactified, and (barring the appearance of singularities, coordinate or otherwise), the full Cauchy development should by revealed in finite $t$. It is also straightforward to see that any causal curve must be interior to the radial lightcones $dt=\pm dr$, and thus this coordinate system automatically gives us a Penrose compactification of solutions. Both $r=0$ and $r=\ell\pi/2$ are timelike curves, hence we need boundary conditions for a well-posed Cauchy evolution; at the origin we impose regularity, and at timelike infinity that the metric is AdS with no incoming radiation (Dirichlet conditions on the matter). These boundary conditions are written down explicitly in Appendix \ref{sec:EOM}.  

To source dynamics in the spacetime, we couple the Einstein equations (\ref{efe}) to a complex scalar field $\Psi(t,r,\theta)$, satisfying the Klein-Gordon equation
\begin{equation}\label{kg}
\Box\Psi=0,
\end{equation}
with stress-energy tensor
\begin{equation} \label{eq:stress_energy}
T_{\mu \nu}(t, r) = \frac{1}{2} \Big( \partial_{\mu} \Psi^* \partial_{\nu} \Psi + \partial_{\mu} \Psi \partial_{\nu} \Psi^* - g_{\mu \nu} g^{\rho \sigma} \partial_\rho \Psi \partial_\sigma \Psi^* \Big),
\end{equation}
where an asterisk denotes complex conjugation.
To allow the scalar field to carry angular momentum, yet maintain a circularly symmetric stress-energy tensor, we impose the following ansatz for the scalar field profile (this is the $m=1$ case in Jalmuzna and Gundlach \cite{jalmuzna2017}):
\begin{equation} \label{eq:jg_matter_ansatz}
\Psi(t, r, \theta) = e^{i \theta} \sin(r/\ell) \big[ \phi(t,r) + i \psi(t,r) \big].
\end{equation}
The net, conserved angular momentum of matter is
\begin{equation} \label{J_net_def}
J_{net} = -4 \int_\Gamma T_{\mu\nu} \xi^\mu n^\nu \sqrt{h} \ d^2 x,
\end{equation}
where the integral is performed over a spacelike hypersurface $\Gamma$ with unit timelike normal vector $n^\nu=[(\partial/\partial t)^\nu-\beta(\partial/\partial \theta)^\nu]/\sqrt{f}$, axial Killing vector $\xi^\nu=(\partial/\partial \theta)^\nu$, and induced metric determinant $h=f \rb^2$~\cite{choptuik2004}\footnote{We have a factor of $4$ difference compared to the equivalent expression in~\cite{choptuik2004} due to a different normalization of our scalar field and different normalization of Newton's constant $G$.}. From (\ref{J_net_def}) we can define an angular momentum density
\begin{equation} \label{Jp_def}
J'(t,r) = - 8\pi T_{\mu\nu} \xi^\mu n^\nu \sqrt{h}.
\end{equation}
Using the Einstein equations we can reexpress (\ref{Jp_def}) in terms of the metric only, giving the following expression for the net angular momentum within a disk of radius $r$ at some time $t$:
\begin{equation} \label{J_def}
J(t,r) = \frac{\rb^3 \beta'}{f}.
\end{equation}
For the vacuum ($T_{\mu\nu}=0$) BTZ spacetimes (\ref{J_def}) evaluates to the corresponding (constant) angular momentum of the BTZ black hole~\cite{jalmuzna2017}.

The explicit form of the Einstein (\ref{efe}, \ref{eq:stress_energy}) and Klein-Gordon (\ref{kg}) equations in terms of our metric (\ref{eq:jg_metric}) and scalar field (\ref{eq:jg_matter_ansatz}) ansatz are given in Appendix \ref{sec:EOM}.

\subsection{Diagnostics}
To help interpret aspects of the geometry, we integrate various sets of timelike and null geodesics. Owing to the circular symmetry of the spacetime, each geodesic possesses a conserved angular momentum $L$; we have only investigated $L=0$ geodesics here.  We also compute a few other diagnostic quantities.  Among these are the Ricci scalar ($R \equiv R^{\mu}_{~\mu}$) and Kretschmann scalar ($K \equiv R^{\mu \nu \rho \sigma} R_{\mu \nu \rho \sigma}$), constructed from the Riemann curvature tensor $R_{\mu \nu \rho \sigma}$.  We also compute the Hawking quasilocal mass aspect $M_H \equiv \frac{\bar{r}^2}{\ell^2} - (\nabla \bar{r})^2$, which in vacuum equals the BTZ mass $M$ in the limit $r\rightarrow \frac{\ell \pi}{2}$. Throughout our evolution we monitor the outgoing null expansion $\Theta = (\partial_t + \partial_r) \bar{r}$, and keep track of the corresponding horizons where $\Theta=0$. In our dynamical spacetimes $M_H$ and $J$ (\ref{J_def}) asymptote ($r\rightarrow \frac{\ell \pi}{2}$) to the conserved mass and angular momentum of the spacetime; when a black hole forms, at late times these values converge to quantities consistent with the proper circumference of the corresponding BTZ black hole's event horizon (\ref{eq:rbar_pm}), which we measure on the outermost apparent horizon (outermost marginally trapped surface).

\section{Numerical Methods} \label{sec:numerical_method}
We follow the methods of~\cite{pretorius2000,jalmuzna2017} to evolve the Einstein-Klein-Gordon system outlined above and in Appendix~\ref{sec:EOM}. In brief, we use a so-called free evolution scheme. Here, the constraint equations are only solved at $t = 0$, after which the Einstein evolution and Klein-Gordon equations are used to evolve all metric and scalar quantities forward in time. The constraints are monitored during evolution, and their convergence to zero checked to ensure we have a self-consistent solution (see Appendix~\ref{sec:convergence_tests}). We adopt a Crank-Nicolson finite difference scheme that is second order accurate in time, along with fourth order spatial differences in $r$.  We also implement Kreiss-Oliger style dissipation \cite{kreiss1973}, which significantly improves the stability of our algorithm near the outer spatial boundary. We solve the finite difference equations using Gauss-Seidel relaxation, with typical resolutions of up to 8193 gridpoints in $r$, and a Courant factor $\lambda \equiv \frac{\Delta t}{\Delta r} = 0.1$.

\subsection{Singularity excision}\label{sec:excision}

The novel feature of our numerical method is our excision procedure, which improves upon that of \cite{pretorius2000,jalmuzna2017} in that it allows us to evolve the spacetime significantly beyond the formation of an apparent horizon. Unlike studies of black hole exteriors, where the purpose of excision is to remove the interior singularities from the domain to allow long-term evolution of the exterior, here we are of course very much interested in uncovering as much of the interior as possible. Therefore, we choose for our excision criterion growth of the magnitude of metric variables above a certain threshold (specifically, $|\dot{A}|\geq 300$ and/or $|\dot{B}|\geq 300$), above which a divergence is usually imminent. When the threshold is reached at a given point, we excise it and all points to its causal future. This procedure results in an excision boundary which is locally only null or spacelike, with the latter occurring if our threshold criterion is satisfied at multiple spacelike separated points. Such an excision boundary makes physical sense, and is necessary for a mathematically well-posed problem. Physically, if somehow a timelike singularity formed, then to reveal it would require a prescription to ``resolve'' the singularity to allow evolution of the spacetime within its lightcone, which the Einstein equations cannot provide. Mathematically then, we cannot place boundary conditions on the excision surface.  Numerically, this can only be stably implemented if no physical characteristics of the equations point into the computational domain from the excised region; such behavior is guaranteed by causality if the excision boundary is spacelike or null. To ascertain the nature of the excision boundary in a given scenario in the continuum limit, we do convergence studies by both increasing the grid resolution and raising our excision threshold criterion, then extrapolating relevant diagnostic quantities (such as curvature scalars or matter energy density) to an extrapolated continuum limit of the excision surface.

\begin{figure}[]
	\centering
	\includegraphics[width=\columnwidth]{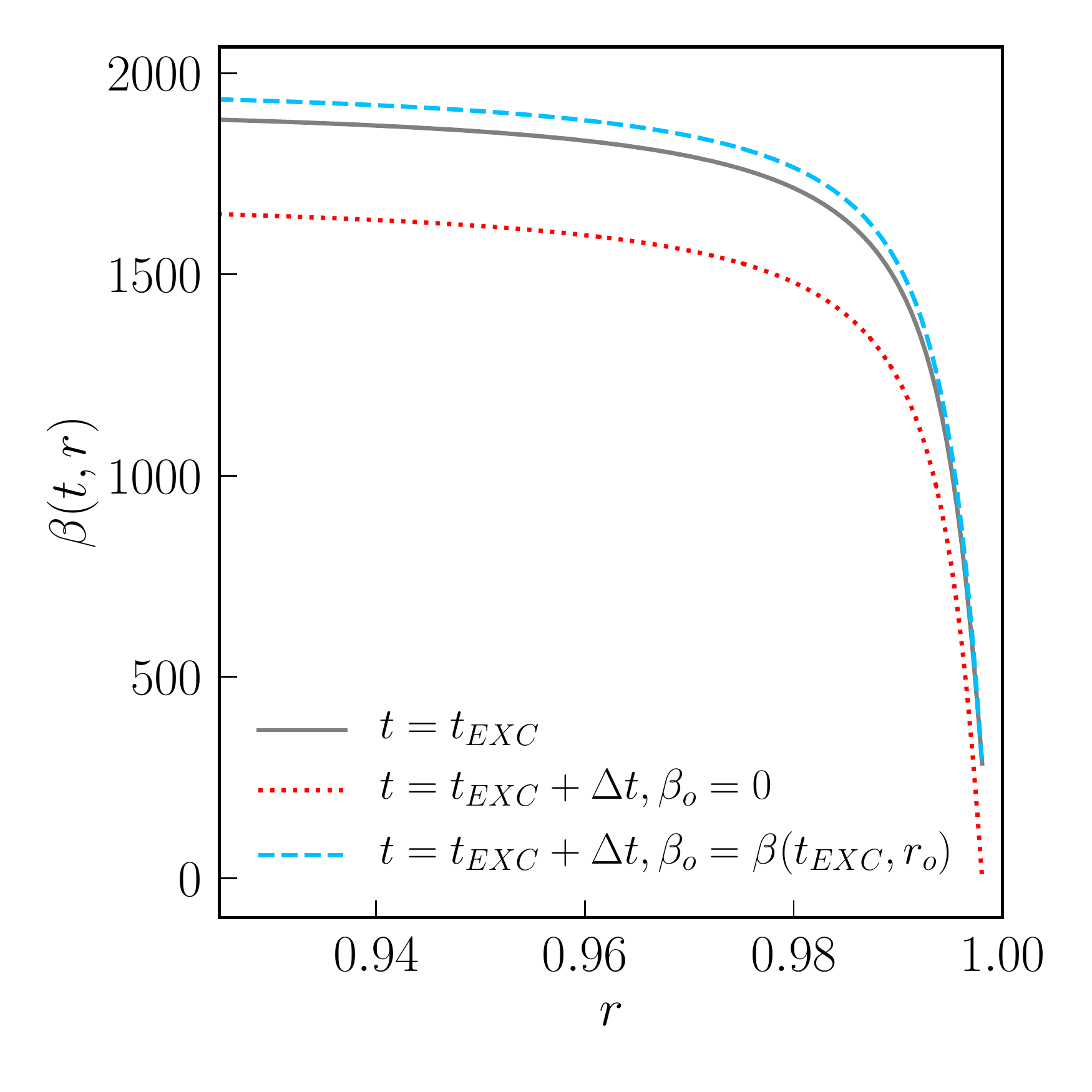}
	\caption{Plot of the metric quantity $\beta(t,r)$ at the first timestep where a portion of the grid is excised ($t = t_{EXC}$; solid gray line), along with $\beta$ one timestep later for the two choices of $\beta_o$ mentioned in Sec. \ref{sec:excision}.  Note that the case with $\beta_o = 0$ (dotted red line) introduces a significant jump in time at the outermost nonexcised gridpoint $r_o$, while the choice $\beta_o(t+\Delta t, r_o) = \beta(t, r_o)$ (dashed blue line) does not.  We adopt the latter choice for $\beta_o$, as it improves the stability of our algorithm.}
	\label{fig:beta_after_excision}
\end{figure}

One technical difficulty in implementing excision within our coordinate system is integrating our evolved variable $\gamma$ to find the metric variable $\beta$. In the reduction of the equations to first order form (see Appendix \ref{sec:EOM}), we define $\gamma(t,r) \equiv \beta'(t,r)$, and evolve a function of $\gamma$ forward in time using the Einstein equations.  Specifically, we evolve $J\equiv \rb^3 \gamma/f$ (\ref{J_def}) via 
(\ref{eq:J_evol_eqn}), then after each timestep compute $\beta$ using
\begin{equation} \label{eq:beta_definite}
\beta(t, r) = - \int_{r_o}^{r} \gamma(t, \tilde{r}) d\tilde{r} \ \ + \ \beta_o(t,r_o).
\end{equation}
Hence we integrate from a larger radius $r_o$ inward, and $\beta_o$ is essentially an arbitrary function of time representing residual gauge freedom in our choice of angular coordinate $\theta$. When the outer boundary is at timelike infinity, $r_o=\ell\pi/2$, we require $\beta_o(t,\ell\pi/2)=0$ for regularity (which is why we integrate from large radii inward, as it makes it easy to enforce this condition). The difficulty with excision comes in when the event horizon curve $\rb_+(t,r)$ reaches the outer boundary, and we then need to excise inward along the Cauchy horizon, so $r_o < \ell\pi/2$. At first, we imposed what we thought would be the simplest choice for the integration constant along the Cauchy horizon, $\beta_o(t,r_o)=0$. However, $\beta(t,r)$ tends to grow very rapidly moving inward in $r$ just prior to excision (see Fig. \ref{fig:beta_after_excision}), so setting $\beta_o(t,r_o)=0$ on the ingoing excision surface introduces significant gauge dynamics that make it challenging to achieve convergent results. Instead then, at each timestep $t+\Delta t$, we set $\beta_o(t+\Delta t,r_o)=\beta(t,r_o$); this procedure essentially freezes $\beta$ at a given point $r_o$ on the excision surface to the value it had at the most recent time when $r=r_o$ was in the interior of the computational domain.

\section{Results} \label{sec:results}

\begin{figure*}[]
	\centering
	\includegraphics[width=\textwidth]{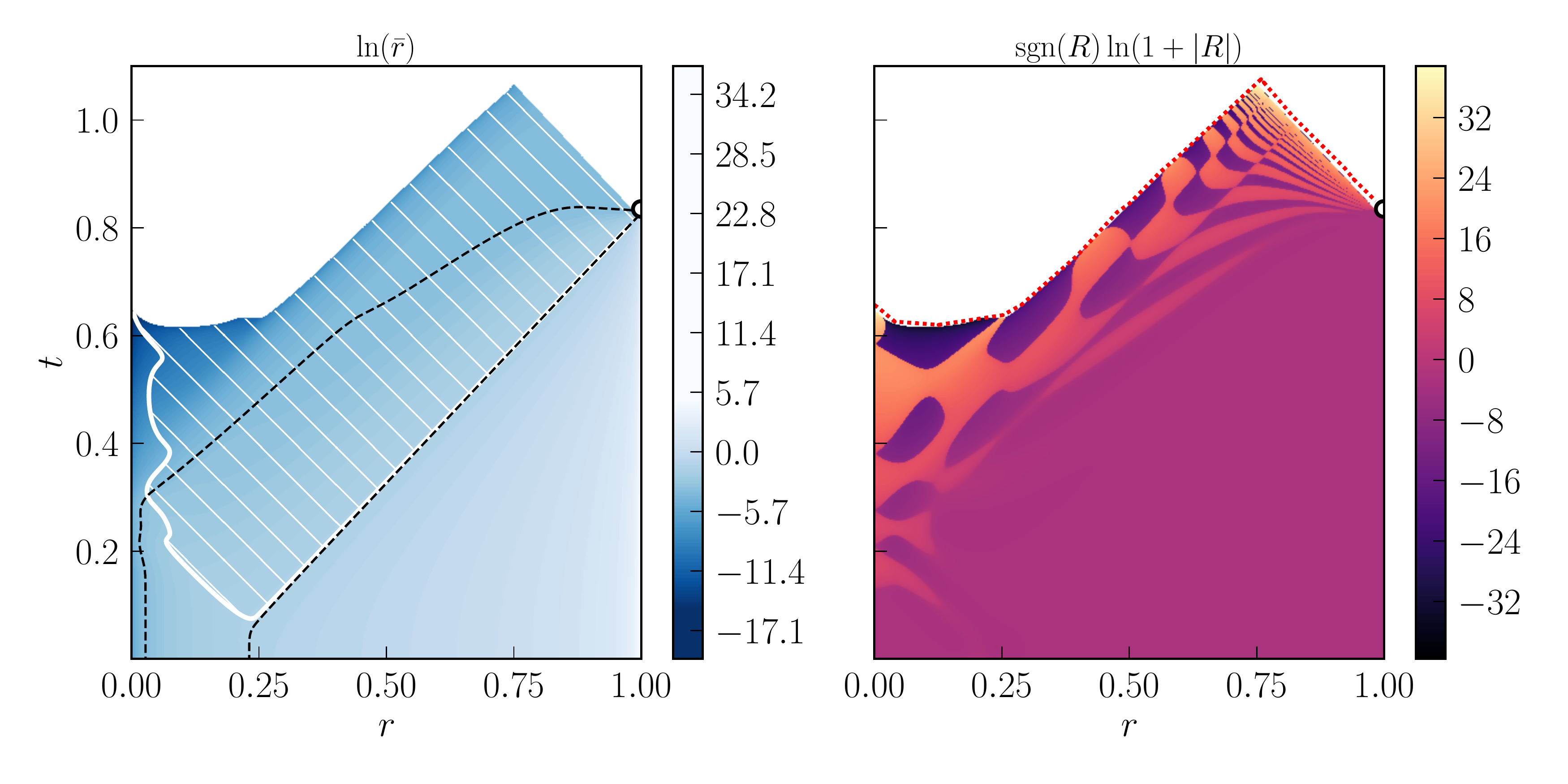}
	\caption{Penrose diagrams from an evolution with spin parameter $a = 0.22$.  Left panel: the color contours correspond to $\ln(\bar{r})$. The thick solid white contour indicates the location where the null expansion $\Theta = 0$, and cross hatching (thin diagonal white lines) denotes the trapped region $\Theta < 0$. For reference, the two contours $\rb=\rb_\pm$ are highlighted by dashed black lines, and correspond to the proper circumferences of the inner and outer horizons of a BTZ black hole (\ref{eq:rbar_pm}) with the same mass $M$ and spin $a$ as that of this spacetime.  Right panel: the color contours depict the Ricci scalar $R$---specifically a signed function that goes like $\pm \ln(R)$ at large $R$, and linearly interpolates between the two branches near $R = 0$.  In both panels, the solid white region is excised from the grid during numerical evolution as explained in Sec.~\ref{sec:excision}.  The red dotted line along the spacelike and null parts of the excision boundary on the right panel indicates the surface where curvature is infinite (measured by both $R$ and the Kretschmann scalar $K$), as determined by extrapolation (see Sec. \ref{sec:CH}). The black circle is the location of future timelike infinity $i^+$, also determined by extrapolation. (Note that there is a piece of the spacelike excision surface that looks almost outgoing null, however our extrapolation to the singularity still gives a spacelike surface here.)}
	\label{fig:a2_contour_plots}
\end{figure*}

\begin{figure*}[]
	\centering
	\includegraphics[width=\textwidth]{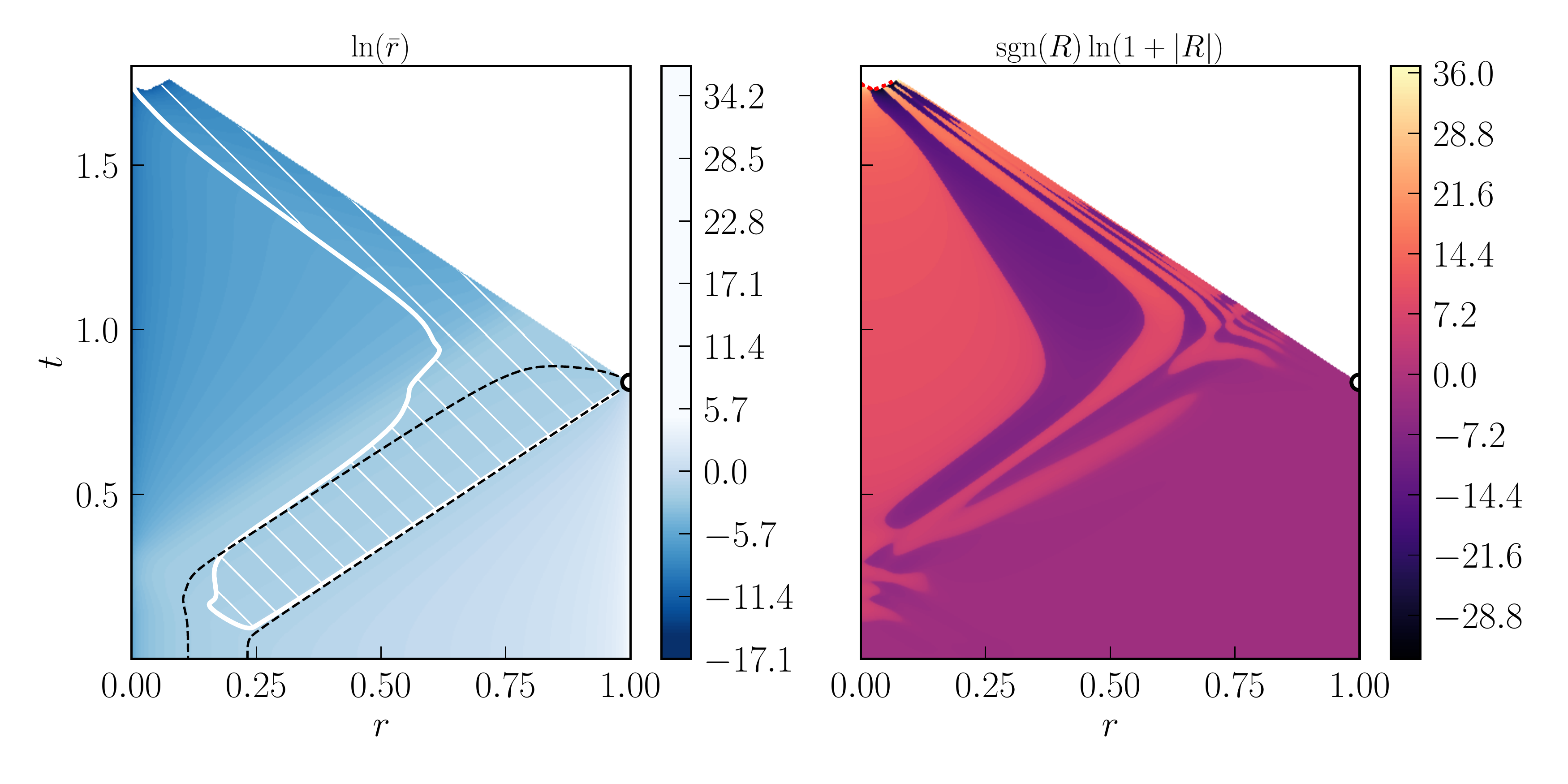}
	\caption{Contour plots analogous to those of Fig. \ref{fig:a2_contour_plots}, for an evolution with $a =0.77$.  Note that the size of the spacelike branch in coordinate $r$ has decreased significantly, and that the dotted red line on the right panel indicating a surface of infinite curvature (as measured by $R, K$) is only present beyond this small spacelike branch.}
	\label{fig:a77_contour_plots}
\end{figure*}

\begin{figure*}[]
	\centering
	\includegraphics[width=\textwidth]{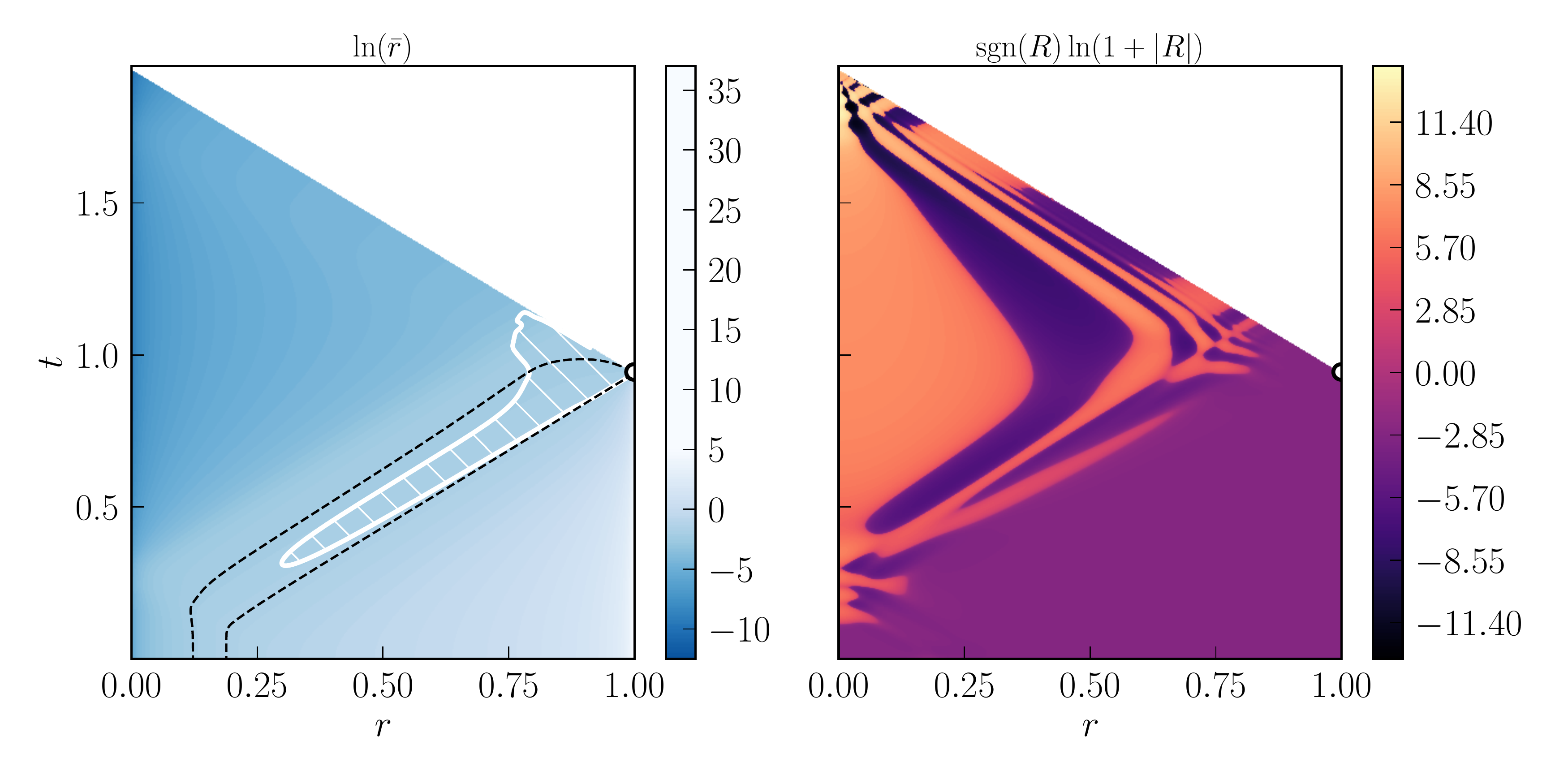}
	\caption{Analogous plot to Fig. \ref{fig:a2_contour_plots}, except this time with $a = 0.91$.  Note the absence of a spacelike excised branch, as well as the absence of surfaces of infinite curvature (as measured by $R, K$) on the right panel.}
	\label{fig:a9_contour_plots}
\end{figure*}

In this section we present our results, beginning in Sec. \ref{sec:id} with a description of the initial data we use. We have explored other classes of initial data and a large range of parameters, though for clarity of the discussion we focus on three particular cases that are representative of the three qualitatively different interior structures we have found, as depicted in the right three panels of Fig.~\ref{fig:a_dep_penrose_diagrams}. Detailed descriptions of the solutions for these three cases are given in Sec.~\ref{sec:bh_interior}.

\subsection{Initial data}\label{sec:id}
We follow the same procedure as~\cite{jalmuzna2017} to construct our initial data. Of the scalar field variables, the quantities $\phi$, $\psi$, $V\equiv \dot{\phi}+\beta\psi$ and $W\equiv \dot{\psi}-\beta\phi$, are freely specifiable at $t=0$. For our metric variables, at $t=0$ we can choose $B=\dot{B}=0$; $A,\dot{A}$ and $\beta$ (via $\gamma$) are then constrained via Eqs. (\ref{eq:Hamiltonian_constraint}), (\ref{eq:Momentum_constraint}) and (\ref{eq:J_constraint}) respectively. 

Here we investigate evolution of the following family of approximately ingoing Gaussian pulses for the scalar field: first defining 
\begin{equation}
F(r,A_0,r_0,\sigma)=A_0 e^{-(r-r_0)^2/\sigma^2},
\end{equation}
where $A_0,r_0$ and $\sigma$ are constant parameters, we choose
\begin{eqnarray}
\phi(t=0,r)&=&F(r,A_0,r_{0\phi},\sigma)+F(-r,A_0,r_{0\phi},\sigma),\ \ \ \ \nonumber \\
V(t=0,r)&=&F'(r,A_0,r_{0\phi},\sigma)+ F'(-r,A_0,r_{0\phi},\sigma),\ \ \ \ \nonumber \\
\psi(t=0,r)&=&F(r,A_0,r_{0\psi},\sigma)+F(-r,A_0,r_{0\psi},\sigma),\ \ \ \ \nonumber \\
W(t=0,r)&=&F'(r,A_0,r_{0\psi},\sigma)+ F'(-r,A_0,r_{0\psi},\sigma).\ \ \ \
\end{eqnarray}
Superposing a Gaussian with its reflection about $r=0$ is a simple way to ensure regularity of the field at $r=0$. There are numerous ways to provide angular momentum in the initial data (see the source term $S_{\gamma'}$ in (\ref{eq:matter_source_terms})); in the above it comes from the Gaussians for $\phi$ and $\psi$ being centered at different locations ($r_{0\phi}$ and $r_{0\psi}$ respectively). We quantify the amount of spin in terms of the BTZ spin parameter $a \equiv |J| / (M \ell) \in [0, 1]$, and we study the formation of black holes with spins ranging from $a = 0.2$ to $a = 0.97$.  For $a < 0.2$ the qualitative structure of the black hole is similar to that of the $a=0.22$ case, but the size of the spin-dependent features, in particular the Cauchy horizon, shrinks as $a \to 0$, presumably smoothly connecting to the $a = 0$ case (see \cite{pretorius2000}).  Thus, we focus on higher spins where we can clearly resolve all the interior features.  For $0.97 < a < 1$ our method breaks down at the numerical resolutions we are currently able to achieve. We have checked a couple of cases where the initial data has $a>1$, and no black hole (or any singular structure) formed within the time it took for the pulse of scalar field to traverse the universe several times.

In the remainder of this section, we present results from three specific cases, $a\in(0.22,0.77,0.91)$, that are representative of the qualitatively different interiors we observe, as illustrated in Fig.~\ref{fig:a_dep_penrose_diagrams} above. The particular initial data parameters are $\sigma = 0.05$ for all cases, and: $r_{0\phi} = 0.225, r_{0\psi} = 0.232, A_{0} = 0.28$ for $a = 0.22$; $r_{0\phi} = 0.2, r_{0\psi} = 0.25, A_{0} = 0.28$ for $a = 0.77$; $r_{0\phi} = 0.2, r_{0\psi} = 0.25, A_{0} = 0.26$ for $a = 0.91$.

\subsection{The black hole interior for three representative cases}\label{sec:bh_interior}

\subsubsection{Penrose diagrams and trapped regions}
In Figs.~\ref{fig:a2_contour_plots}, \ref{fig:a77_contour_plots}, and \ref{fig:a9_contour_plots} we show proper circumference $\rb$ and Ricci scalar $R$ on Penrose diagrams for the evolutions with $a=0.22$, $0.77$ and $0.91$ respectively. In all cases a trapped region forms soon after evolution begins, though it does so more rapidly for the two lower spin cases. The trapped region first appears at a single nonzero proper circumference, and as it grows is bounded by an outer and inner apparent horizon. The former rather quickly asymptotes to the null event horizon of the spacetime, whose late-time circumference is consistent with that of a BTZ black hole, $\rb_+$ (\ref{eq:rbar_pm}), with the same mass $M$ and angular momentum $J$ as that of the spacetime. For the lowest spin case (Fig.~\ref{fig:a2_contour_plots}), the inner horizon quickly collapses to $\rb=0$, and never resembles an outgoing null surface as in the BTZ spacetime. For the intermediate spin case (Fig.~\ref{fig:a77_contour_plots}), at intermediate times the inner horizon does appear close to null and to $\rb=\rb_-$, though at late times also collapses to $\rb=0$. For the high spin case (Fig.~\ref{fig:a9_contour_plots}), the inner horizon also is almost null at intermediate times, though not at the $\rb=\rb_-$ of the corresponding BTZ spacetime, and eventually runs into the Cauchy horizon (we discuss below in Sec.~\ref{sec:CH} why we identify the ingoing null part of the excised region as the Cauchy horizon, and not merely the causal future of a coordinate singularity). In other words, for the high spin case, there is a large portion of the Cauchy development of the interior that never becomes trapped.

\subsubsection{The Cauchy horizon} \label{sec:CH}

\begin{figure}[h!]
	\includegraphics[height=0.35\textheight]{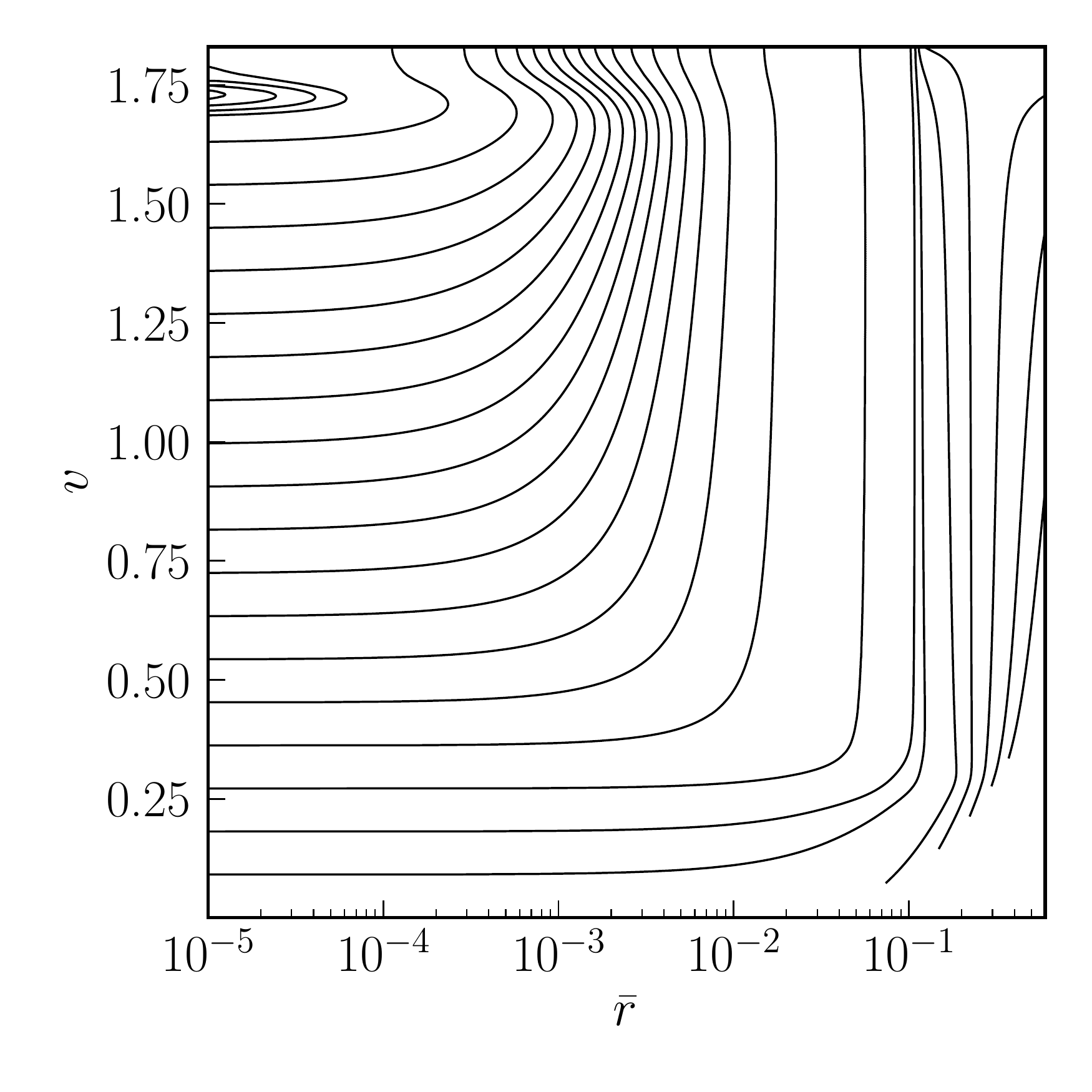}
	\caption{Outgoing null rays plotted as a function of the circumferential radius $\rb$ and a coordinate advanced time $v \equiv t + r$, for the case with $a = 0.77$ (compare to Fig.~\ref{fig:a77_contour_plots}).  Note that there are three groups of null rays: those that begin at the origin and return to the origin, at which time $\rb = 0$ has become a spacelike singularity; those that asymptote to the Cauchy horizon, reaching a finite value of $\rb$ at late times; and those that escape before the black hole forms, eventually reaching the timelike AdS boundary $\rb = \infty$.}
	\label{fig:outgoing_null_plot}
\end{figure}

Figs.~\ref{fig:a2_contour_plots}-\ref{fig:a9_contour_plots} suggest that the extrapolated point where the apparent horizon meets the outer boundary is $i^+$, and that the ingoing null branch of the excision surface emanating from the point on the outer boundary near $i^+$ is asymptoting to the Cauchy horizon. If this is the case, the exterior of the spacetime should be complete in the sense that $i^+$ can only be reached in infinite proper time by any causal curve, and the event horizon only ``reaches'' the corresponding point on the Penrose diagram in infinite affine time (and of course the fact that these disparate limits are at the same location on the diagrams is only an artifact of the Penrose compactification). Moreover, in the interior the opposite should hold: any causal curve should reach the Cauchy horizon in finite affine (proper) time. We performed several checks on the numerical solutions to confirm that this behavior occurs. First, we integrated proper time $\tau$ along $\rb={\rm const.}$ timelike curves exterior to the horizon (note that these curves are not geodesics), and extrapolated to the point where $\tau\rightarrow\infty$; these points are shown as the open black circles in Figs.~\ref{fig:a2_contour_plots}-\ref{fig:a9_contour_plots}, and converge to the extrapolated location where the event horizon generator reaches the boundary. We also integrated sets of outgoing null geodesics throughout the spacetime (see Fig.~\ref{fig:outgoing_null_plot} for an example of their trajectories for the $a=0.77$ case), confirming those interior to the event horizon end on the Cauchy horizon or central singularity in finite affine time. 

\subsubsection{Focusing of ingoing geodesics in the interior} \label{sec:OIH}

\begin{figure}[h!]
	\centering
	\includegraphics[width=0.4\textwidth,trim=0.6in 0.0in 0.0in 0.0in]{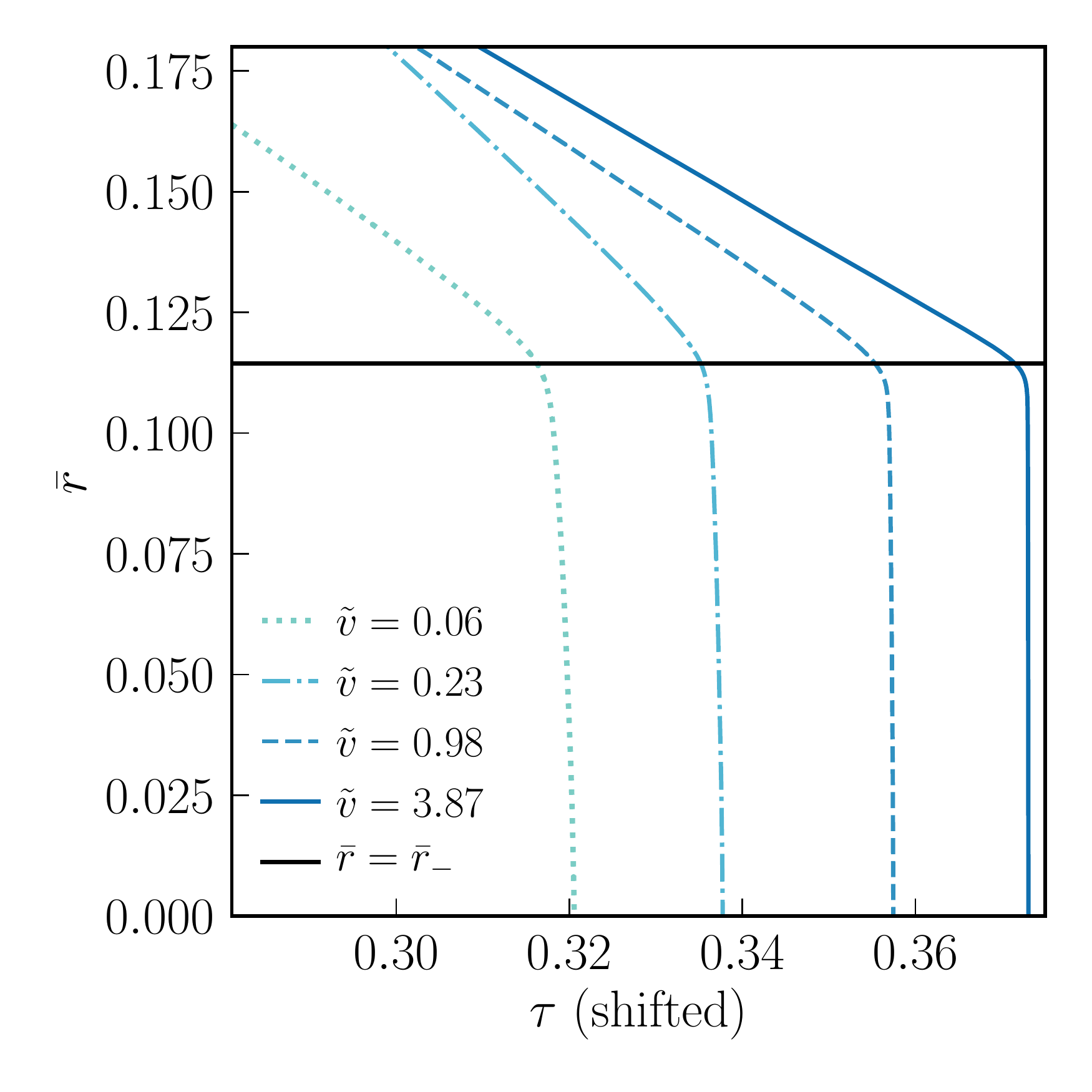}
	\includegraphics[width=0.4\textwidth,trim=0.6in 0.0in 0.0in 0.0in]{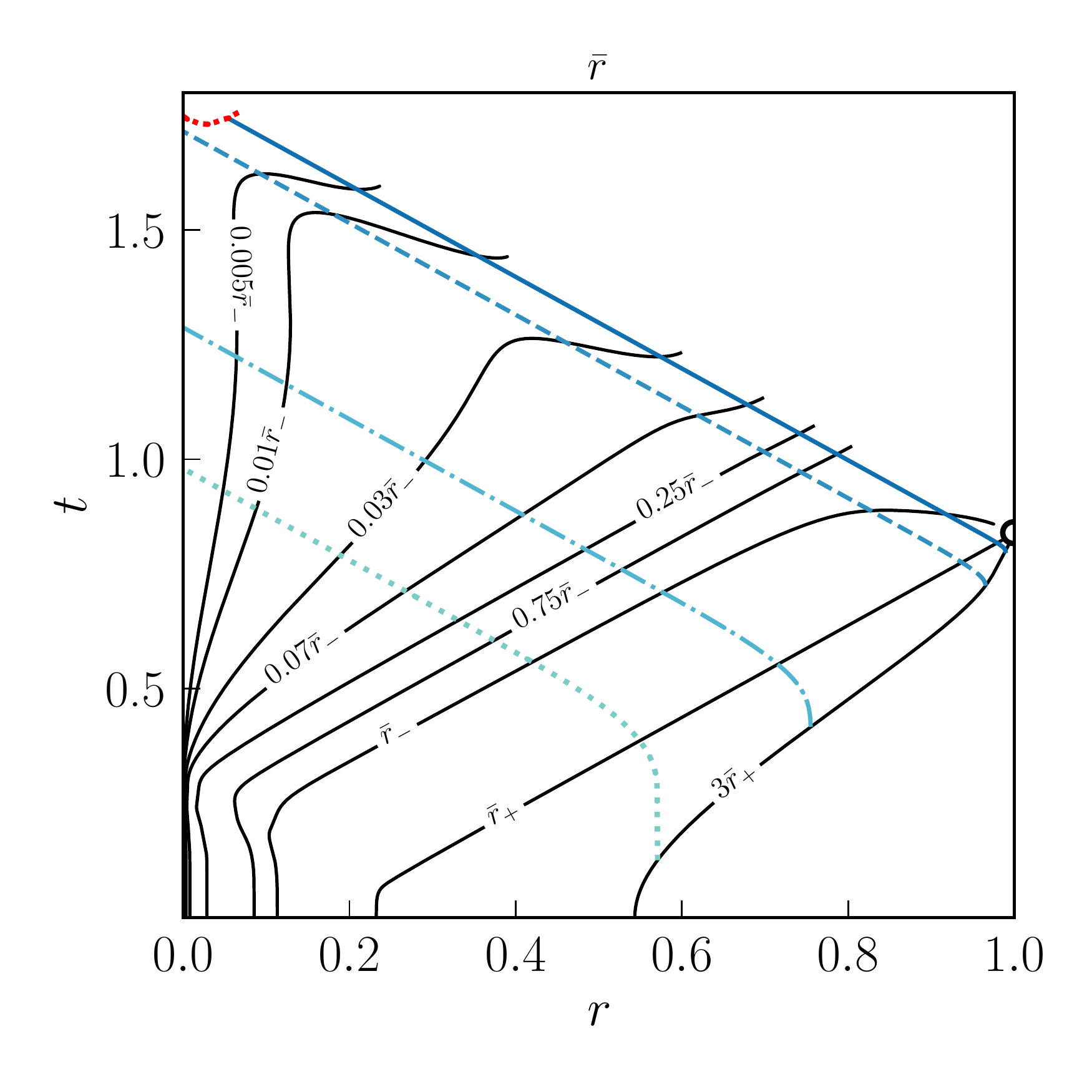}
	\caption{Top: proper circumferences versus proper time $\bar{r}(\tau)$ for several ingoing timelike geodesics, beginning from rest at $\rb=3\rb_+$, at successively later times $\tilde{v}$, for the $a=0.77$ case.  Bottom: the same geodesics on a contour plot of $\bar{r}$, along with $i^+$ and the infinite curvature surface as in Fig.~\ref{fig:a77_contour_plots}. The time parameter $\tilde{v}$ is, as discussed in Sec.~\ref{sec:AdS3_and_BTZ}, the affine time along the outgoing null generator of the event horizon when the given geodesic crosses it. For clarity in the top figure a constant shift has been added to $\tau$ for each geodesic to display them in increasing order in $\tilde{v}$, and the horizontal solid line is the location of the inner horizon if the spacetime were BTZ, $\rb=\rb_-$ (which is {\em not} the location of the inner horizon in the dynamical spacetime; see Fig.~\ref{fig:a77_contour_plots}). Note that the geodesics are strongly focused to the origin beyond $\bar{r} = \bar{r}_-$, and the sharpness of the focusing increases with $\tilde{v}$. The same phenomenon occurs for null geodesics in terms of their affine parameters; see Fig.~\ref{fig:geodesic_sharpening} for measurements of the sharpening rate for both classes of geodesic.}
	\label{fig:OIH_knee}
\end{figure}

\begin{figure}[h!]
	\centering
	\includegraphics[width=0.45\textwidth]{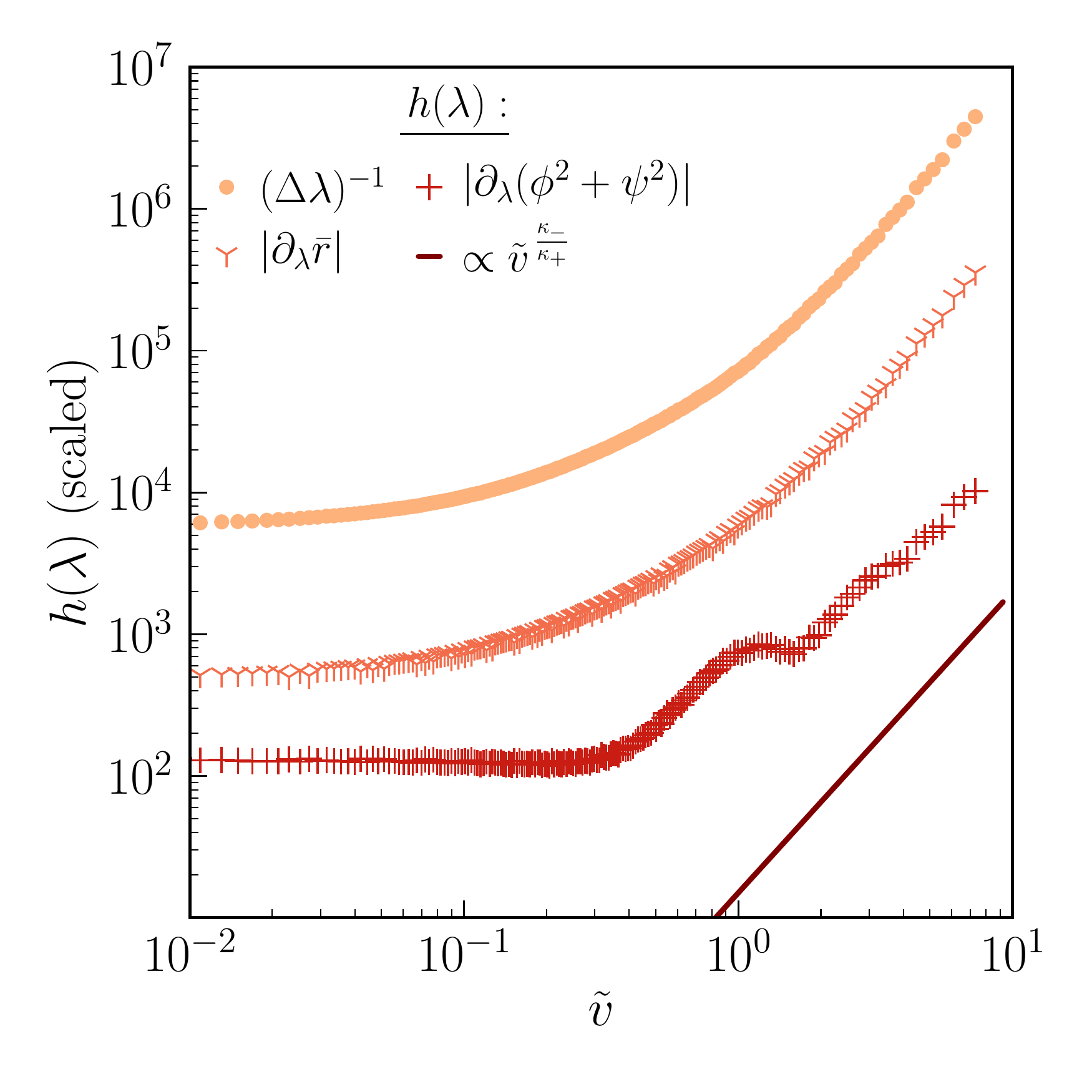}
	\includegraphics[width=0.45\textwidth]{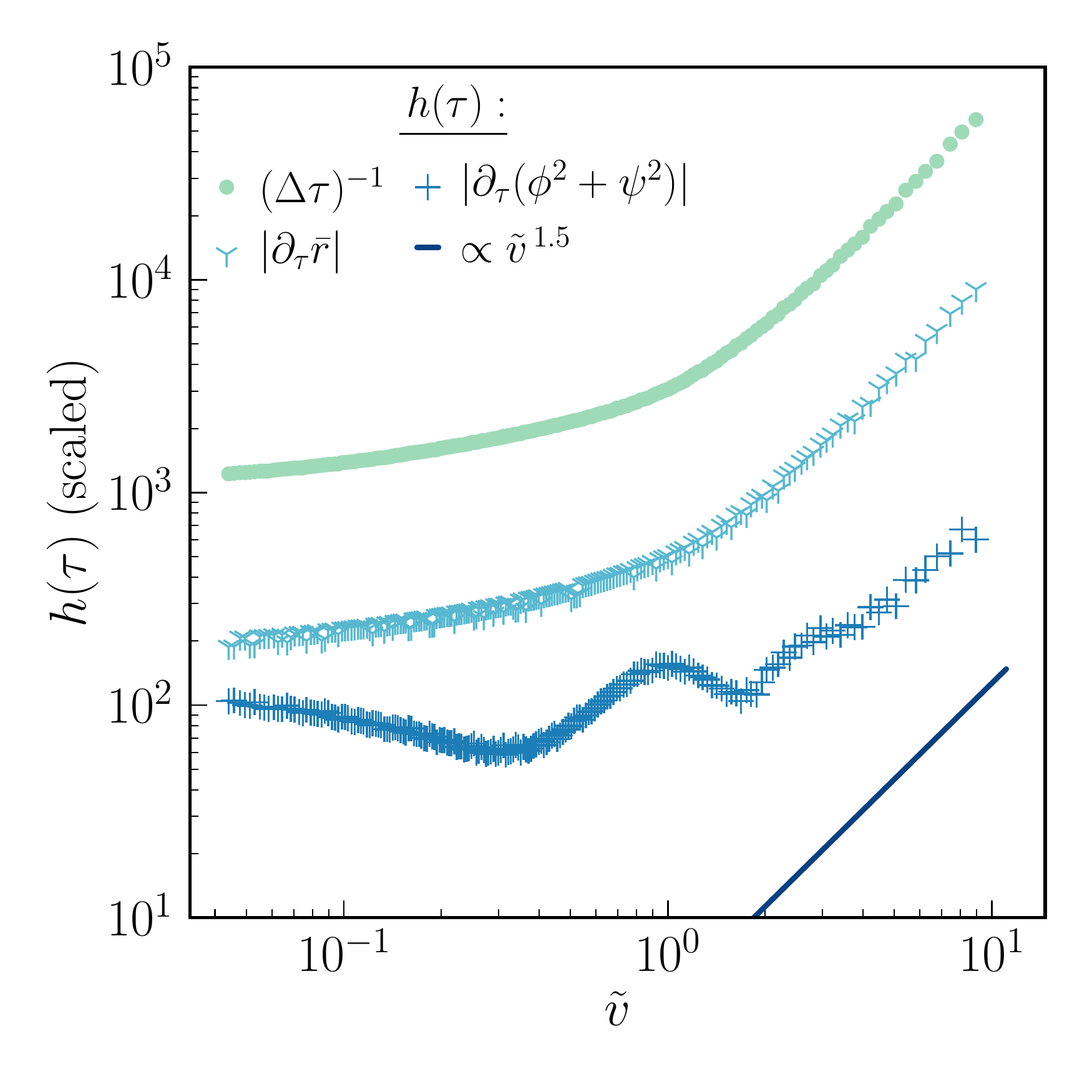}
	\caption{On the top (bottom) panel, sharpening of various quantities along ingoing null (timelike) geodesics in terms of their affine parameters $\lambda$ (proper times $\tau$), as a function of the timing coordinate $\tilde{v}$ (see Sec. \ref{sec:AdS3_and_BTZ} and the caption of Fig. \ref{fig:OIH_knee}), from an evolution with $a = 0.77$. The gradients of $\rb$ and square of the magnitude of the scalar field $\phi^2+\psi^2$ are measured when the corresponding geodesic crosses $\rb=0.75 \rb_-$, and the net elapsed affine time $\Delta \lambda$ (proper time $\Delta \tau$) is counted between the crossing at $\rb=0.75 \rb_-$ and $\rb=0.25 \rb_-$ (see Fig.~\ref{fig:OIH_knee}). The measured slopes do not depend much on the particular values chosen for the crossing radii, as long as they are past the knee at $\rb\sim\rb_-$ (Fig.~\ref{fig:OIH_knee}).}
	\label{fig:geodesic_sharpening}
\end{figure}

Given how different the inner horizon structures are from their vacuum BTZ black hole counterparts, one might expect the Marolf and Ori focusing effect discussed in Sec.~\ref{sec:AdS3_and_BTZ} to be significantly lessened, or even absent. However, in all three cases $(a=0.22,0.77,0.91)$, at late times approaching the Cauchy horizon, we find a {\em quantitatively} similar growth in the rate of change of interior features as experienced by infalling observers. In Fig.~\ref{fig:OIH_knee} we show proper circumference versus proper time $\rb(\tau)$ for several infalling timelike observers, beginning from rest at $\rb=3 \rb_+$, for the $a=0.77$ case. Notice the near steplike drop in proper circumference that occurs near $\rb_-$. Fig.~\ref{fig:geodesic_sharpening} shows the rate of growth of this feature measured a couple of ways (one following the BTZ calculation outlined in Sec.~\ref{sec:AdS3_and_BTZ}), as well as the change in the observed scalar field, for both ingoing timelike and null geodesics (again for the $a=0.77$ case). At late times the rate of steepening for the null geodesics follows the power law prediction (\ref{null_steep}) quite closely; we have not derived the analogous result for timelike geodesics, though the numerical data also shows a power law, but with a different slope than in the null case. Extrapolating these curves to the Cauchy horizon ($\tilde{v}\rightarrow \infty)$ suggests the knee becomes an actual step function there. In the notation of \cite{2013CMaPh.323...35K} (see also~\cite{VandeMoortel:2019ike,VandeMoortel:2020olr}), this behavior implies that the Cauchy horizon is composed of two null segments, the initial piece $\mathcal{CH}_{i^+}$ emanating from (but not including) $i^+$, where $\rb$ is nonzero except possibly at its future endpoint, connected to $\mathcal{S}_{i^+}$, on which $\rb$ extends continuously to zero.

Such a feature in $\rb(\tau)$ and $\rb(\lambda)$ further implies infalling observers are subject to an asymptotically divergent tidal force, experienced in a region of the Penrose diagram well before any singularity (for the lower $a$ cases) or the Cauchy horizon\footnote{Though ``well before'' is somewhat an artifact of how the region beyond the near-shock feature is magnified on the Penrose diagram; this region is crossed in vanishingly small proper/affine time by geodesics.}. We emphasize that although this feature is encountered as the observers cross $\rb\sim \rb_-$, in the dynamical spacetimes this is {\em not} the location of the inner horizon at late times. It is remarkable and puzzling then that the Marolf-Ori calculation still manages to give the quantitatively correct growth rate, as it seemed to be an essential part of the calculation that $\rb_-$ was marginally trapped---i.e. outgoing geodesics at larger radii have negative expansion, while geodesics at smaller radii have positive expansion, resulting in the localization of features near $\rb_-$. In the dynamical case, the asymptotic scaling regime where the power law rate matches the vacuum calculation is in a region of the spacetime that is fully trapped, and moreover $\rb=\rb_-$ is clearly spacelike there, as is evident from the Penrose diagrams (Figs. \ref{fig:a2_contour_plots}-\ref{fig:a9_contour_plots}).

Though more work is needed to completely understand the geodesic focusing effect, it is worth pointing out a few key features of our solutions which hint at the true cause.  After the ingoing pulse of scalar radiation triggers apparent horizon formation, it passes through the origin and moves outward on a null trajectory, the entire time remaining in a region of spacetime where the null expansion $\Theta$, though negative, is very close to zero.  As a result, the pulse effectively sits at a constant value of $\rb$ (on the nearly null contours of $\rb$ roughly between $0.25 \rb_-$ and $0.75 \rb_-$ in the bottom panel of Fig.~\ref{fig:OIH_knee}) throughout the entire history of the black hole, until it eventually runs into the Cauchy horizon.  This behavior leads us to speculate that the backreaction of this pulse of matter on the geometry gives rise to the large coordinate acceleration the geodesics experience, as shown in the top panel of Fig.~\ref{fig:OIH_knee} (this is consistent with a similar effect calculated in a 4D charged shell model of collapse to a Reissner-Nordstrom black hole~\cite{Garfinkle:2011pj}). It remains unclear, though, why the calculation in the eternal BTZ background (see Sec.~\ref{sec:AdS3_and_BTZ}) correctly predicts the sharpening rate for null geodesics, in particular that the growth is controlled by a number close to the surface gravity of the inner horizon of the vacuum case.

\subsubsection{Regularity of the Cauchy horizon, and presence of a spacelike singularity}

\begin{figure}[h!]
	\centering
	\includegraphics[width=0.45\textwidth,trim=0.6in 0.0in 0.0in 0.0in]{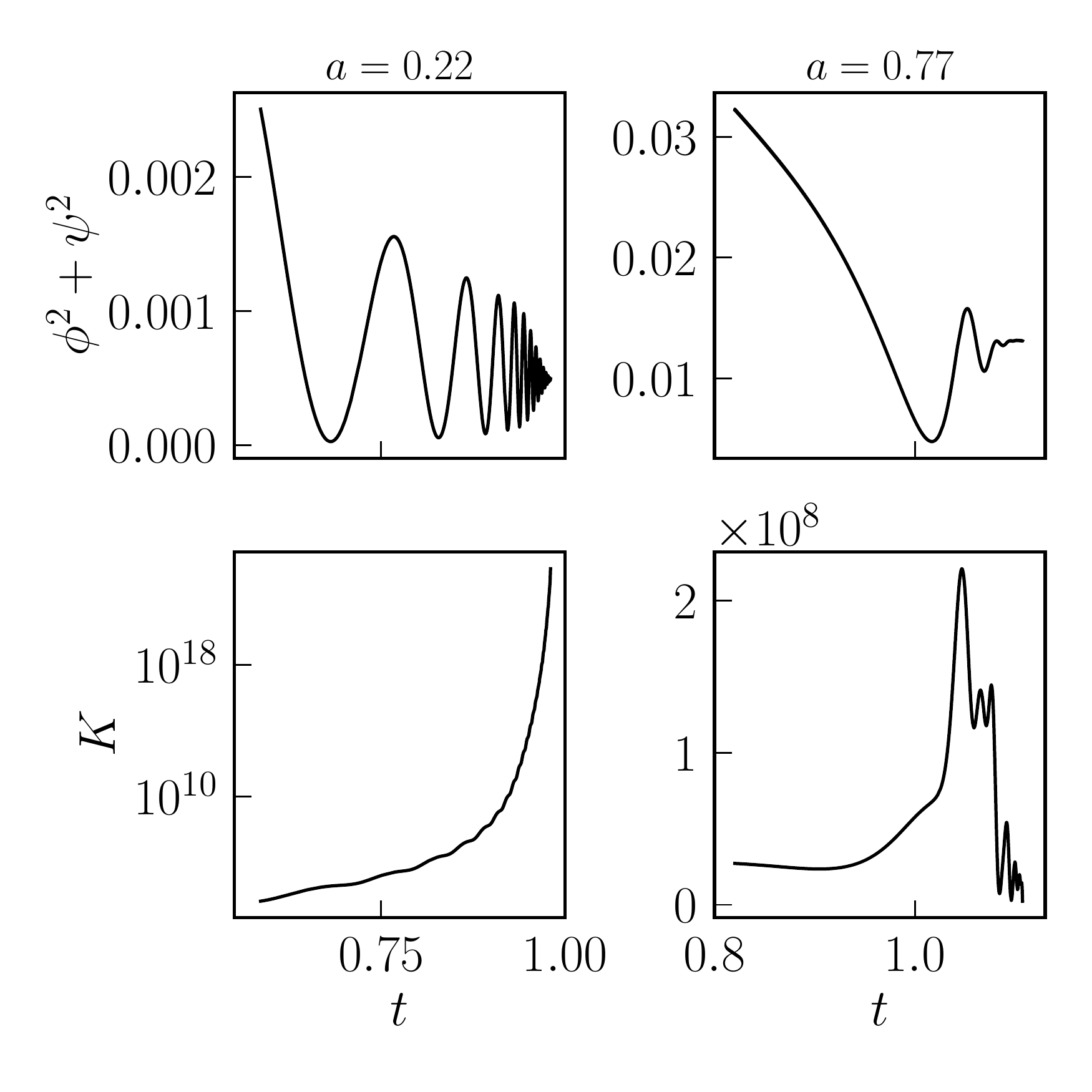}
	\caption{Behavior of the norm of the scalar field components $\phi^2+\psi^2$ and the Kretschmann scalar $K$ for $a = 0.22$ (left column) and $a = 0.77$ (right column) along a representative outgoing null ray approaching the Cauchy horizon.  Note that $\phi^2 + \psi^2$ does not diverge in either case, although its derivative appears to for low spin (upper left panel), sourcing a divergence in the stress-energy and curvature (bottom left panel).  For higher spin (right column) the time derivative of the scalar field remains finite and so does curvature.  For $a = 0.91$, the behavior is qualitatively similar to the $a=0.77$ case shown in the right column.}
	\label{fig:msum_K_CH}
\end{figure}

\begin{figure*}[]
	\centering
	\includegraphics[height=0.3\textheight]{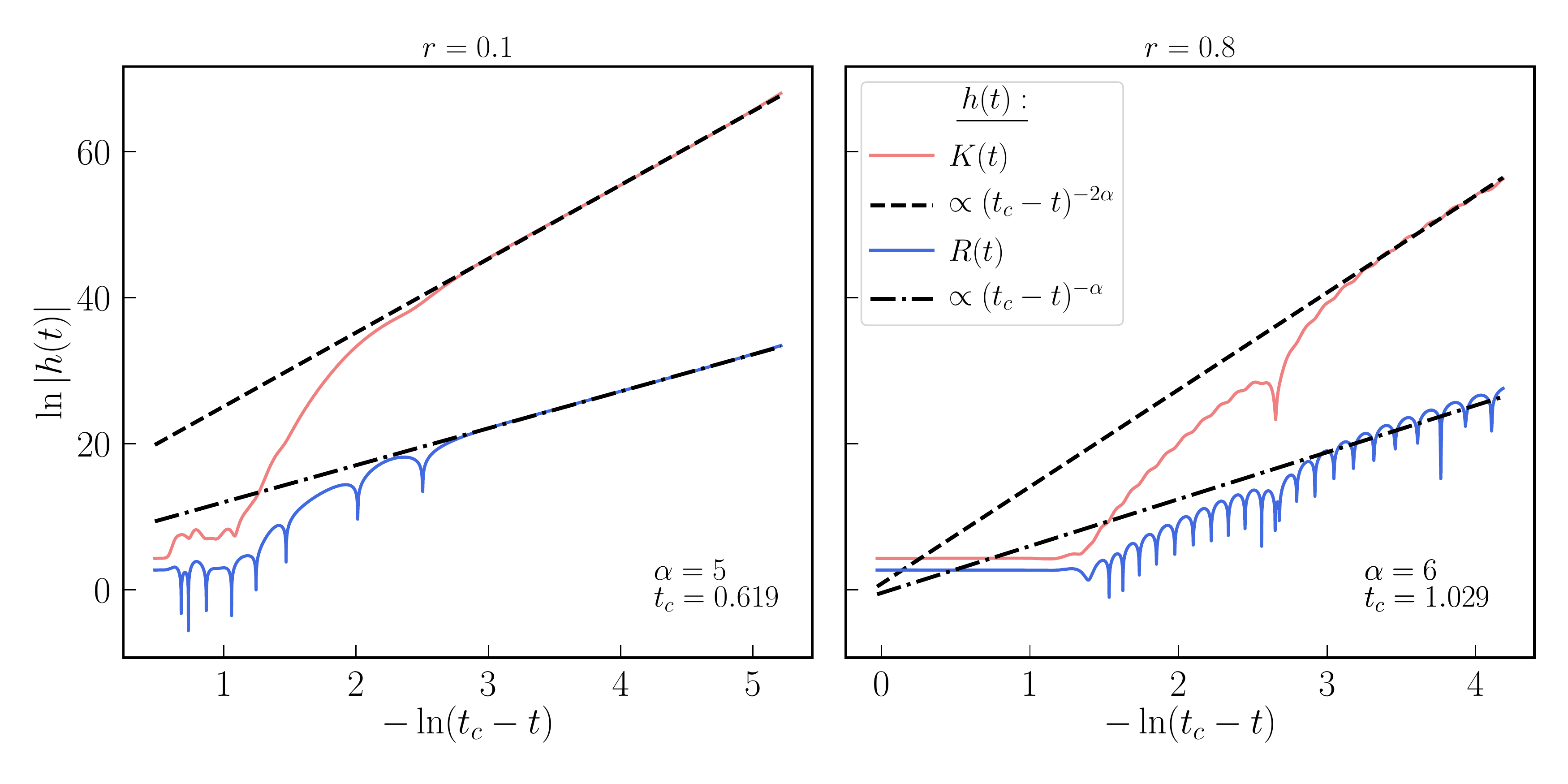}
	\caption{Plots illustrating the method we use to find the presumed surface of infinite curvature, depicted by the dashed red lines on the right panels of Figs.~\ref{fig:a2_contour_plots} and \ref{fig:a77_contour_plots}. Along each $r={\rm const.}$ line, if a quantity $h(t)$ appears to be diverging approaching the excision surface, we assume it does so like $h(t) \propto (t_c - t)^{-\alpha}$, with a constant power $\alpha$ and time-of-divergence $t_c$, and measure the two constants by fitting to the late time behavior as depicted above (this is done independently at each $r$, so in general the ``constants'' vary with $r$). It is important to note that this approach is not based on any theoretical model for the divergence of these quantities; we are merely extrapolating the numerical data. The above examples show the divergence of the Ricci ($R$) and Kretschmann ($K$) curvature invariants approaching the spacelike singularity (left plot) and the singular Cauchy horizon (right plot) for the case with spin $a = 0.22$.  Note that $K$ grows at double the rate $R$ does, as one would expect due to the fact that in 3D they obey the relation $K = 4 R_{\mu \nu} R^{\mu \nu} - R^2$.}
	\label{fig:extrapolation_plot}
\end{figure*}

Dias, Reall, and Santos \cite{dias2019}, through study of linear perturbations of BTZ, found that a massless scalar field should be of differentiability class $C^{\lfloor \underline{b} \rfloor}$ (where $\lfloor \underline{\cdot} \rfloor$ gives the largest integer strictly less than its argument) at the Cauchy horizon, where
\begin{equation}
b = \frac{2}{\frac{\bar{r}_+}{\bar{r}_-} - 1}.
\end{equation}
As a function of $a \equiv |J|/(M \ell)$, one may easily combine the above equation with (\ref{eq:rbar_pm}) to find that the field should be $C^0$ for $a \leq 0.6$, $C^{1}$ for $0.6 < a \leq 0.8$, and increasingly regular for higher spins.  Furthermore, by full contraction of the Einstein equation we have that the Ricci scalar is proportional to the trace of the stress-energy tensor, so the Ricci scalar $R$, and consequently the Kretschmann scalar $K$, should diverge if the scalar field is $C^0$, but not if it is $C^1$ or greater. 

Our fully nonlinear results appear to agree with this linear analysis. In Fig.\,\ref{fig:msum_K_CH} we illustrate the behavior of the scalar field and $K$ along a representative outgoing null geodesic approaching the Cauchy horizon. To determine the infinite curvature surfaces denoted by the dashed red lines on the right panels of Figs.~\ref{fig:a2_contour_plots} and \ref{fig:a77_contour_plots}, we extrapolate the growth of $R$ and $K$ in coordinate time $t$ along coordinate $r={\rm const.}$ lines, as illustrated in Fig. \ref{fig:extrapolation_plot}.  As a secondary check, we also extrapolate along ingoing and outgoing null geodesics; both approaches yield consistent locations for the infinite curvature surface. Of the three cases presented here, only the $a = 0.22$ case shows singular behavior of $R$ and $K$ on the Cauchy horizon. The Hawking mass $M_H$ displays a similar trend, only diverging on the Cauchy horizon for the $a=0.22$ case. In the larger spin cases where curvature is finite, the analysis of \cite{dias2019} suggests there should still be some loss of regularity at the Cauchy horizon; with our second order accurate code and the resolutions we have run, we are not able to extrapolate higher derivatives of the scalar field with enough accuracy to make any definitive statements in this regard.

\begin{figure}[h!]
	\centering
	\includegraphics[width=0.45\textwidth,trim=0.6in 0.0in 0.0in 0.0in]{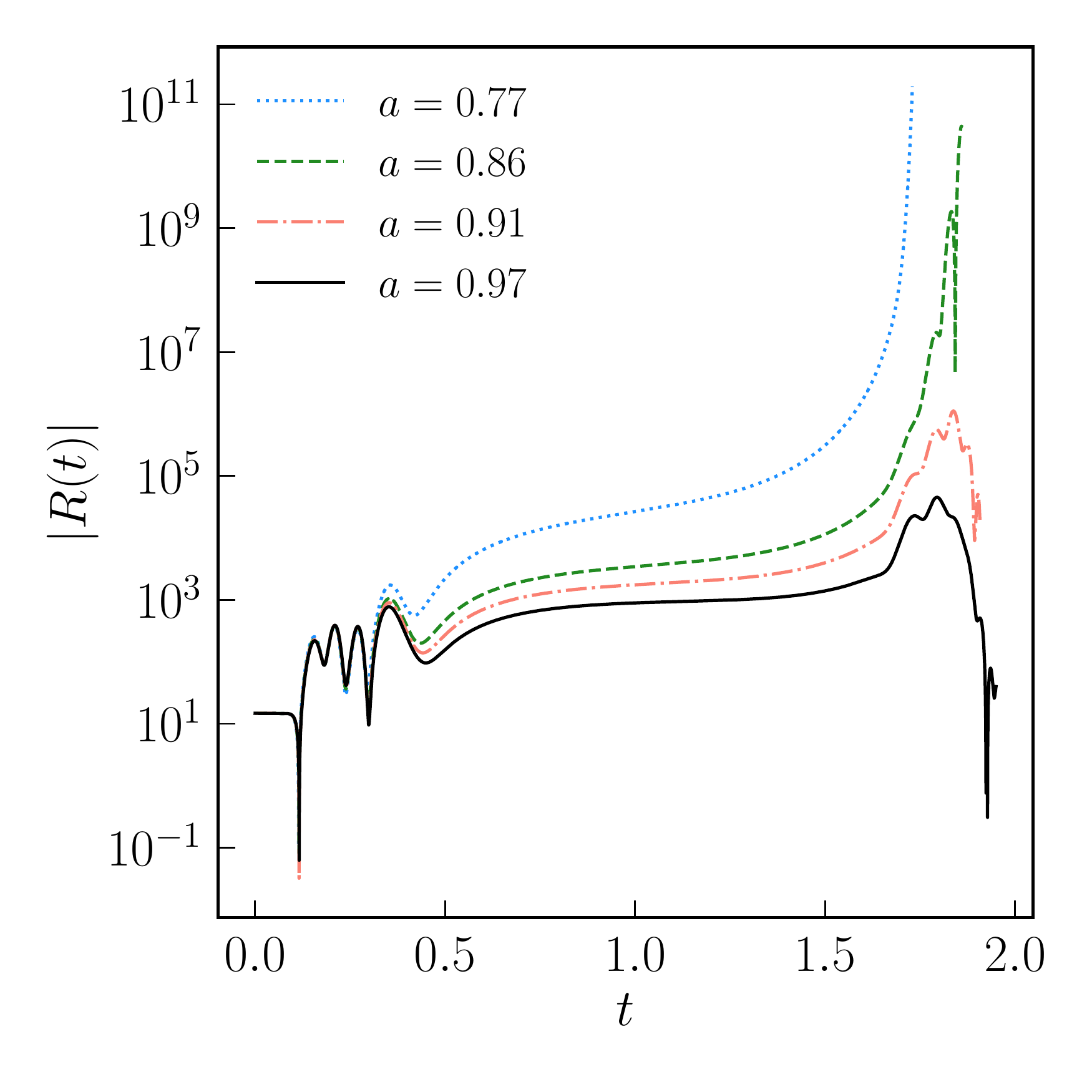}
	\caption{The magnitude of the Ricci scalar $R$ as a function of coordinate time $t$ at $r = 0$ as a function of spin $a$.  We find that for spins greater than $a \approx 0.87$, $R(t)$ remains finite and goes through a local maximum before decreasing in the approach to the Cauchy horizon, whereas for lower spins it trends toward a divergence there.}
	\label{fig:Ricci_origin}
\end{figure}

As evident from the Penrose diagrams of the three cases, Figs. \ref{fig:a2_contour_plots}-\ref{fig:a9_contour_plots}, the relative size of the spacelike branch of the excision surface (which is always singular when present) decreases compared to the size of Cauchy horizon as the spin increases. Particularly interesting is that for spins greater than $a \approx 0.87$ the spacelike branch vanishes, and the then-regular Cauchy horizon extends all the way in from $i^+$ to meet the regular, timelike origin at $\rb=0$. To our knowledge this is the first example of a null Cauchy horizon formed in a collapse scenario that does not ``break down'' (in the language of~\cite{2013CMaPh.323...35K}) to a different class of singularity in the interior\footnote{In the ``two ended'' 4D Reissner-Nordstrom case, examples have been presented where the Cauchy development of small perturbations in the interior leads to a bifurcate null Cauchy horizon with no spacelike singularity \cite{2014CMaPh.332..729D}; ``two-ended'' models have fundamental differences from the ``one-ended'' spacetimes relevant to gravitational collapse, however \cite{VandeMoortel:2019ike}.}. Figure \ref{fig:Ricci_origin} shows the behavior of the Ricci scalar $R$ as a function of time at the origin, and illustrates the qualitative change in late-time dynamics toward smaller curvature with increasing $a$.

\section{Conclusion} \label{sec:conclusion}

We have numerically constructed circularly symmetric solutions to the Einstein-Klein-Gordon equations in asymptotically $AdS_3$ spacetime, which describe the gravitational collapse of a scalar field with angular momentum to form rotating black holes. We have implemented an excision algorithm that appears to be able to reveal (in the continuum limit) the full Cauchy development of a family of initially (approximately) ingoing smooth Gaussian scalar field pulses. Our main findings, summarized schematically in Fig.~\ref{fig:a_dep_penrose_diagrams}, are that there are four qualitatively different geometric structures describing the future boundary of the Cauchy development, and that which one occurs in a particular collapse depends on the spin parameter $a$ of the black hole that forms. For zero spin, the earlier work~\cite{pretorius2000} revealed that a central (proper circumference $\rb=0$), spacelike singularity forms in the interior. For small spins $0<a\lesssim0.60$, we find that a null branch of a Cauchy horizon forms, emanating from future timelike infinity $i^+$ on the Penrose diagram (but not coincident with $i^+$), along which $\rb$ continuously decreases from the event horizon circumference $\rb=\rb_+$ to $\rb=0$, eventually meeting up with a central spacelike singularity. In this case the Cauchy horizon is ``weakly singular'', in that the metric and scalar field are finite there, but their gradients diverge so that the curvature scalars $R$ and $K$ are singular. The Hawking mass also grows here (the ``mass inflation'' phenomenon), seemingly to a divergence as the Cauchy horizon meets the central singularity. For $0.60\lesssim a \lesssim 0.87$ we find a similar Penrose diagram to the lower spin cases, except that the Cauchy horizon never becomes singular, i.e. scalar field gradients and curvature invariants extrapolate to finite values on it. This behavior is consistent with the linear analysis of~\cite{dias2019}, and as they conclude, shows that formation of rapidly rotating black holes in 3D asymptotically AdS spacetime violates the strong cosmic censorship conjecture (though this picture might change in light of quantum effects; see \cite{emparan2020}). For $0.87\lesssim a\lesssim 0.97$ we find that the central spacelike singularity vanishes, and a regular, null Cauchy horizon extends all the way inward to a regular, timelike origin at $\rb=0$. Our algorithm is not able to follow the full development of the interior for near-extremal ($0.97\lesssim a \leq 1$) black holes, and though we do not see any hints of qualitatively new features emerging compared to the highest spin case we can fully resolve, we cannot make definitive statements about this limit.

For all $a>0$ cases we have studied, we find that the focusing effect experienced by infalling geodesic observers crossing the inner horizon $\rb=\rb_-$ of a vacuum BTZ black hole also occurs in the dynamical collapse interiors. Remarkably, the rate of focusing is quantitatively similar to the vacuum case (as originally derived in~\cite{marolf2012}), and moreover it still occurs at roughly the same radius, despite that now $\rb=\rb_-$ is {\em not} an inner horizon (for low spin cases the inner horizon collapses to the origin $\rb=0$ well before the rate of focusing begins to match the background calculation). One dramatic consequence of this focusing is that timelike (null) observers experience a drop in proper circumference from $\rb=\rb_-$ to $\rb=0$ in ever decreasing proper (affine) time the closer to the Cauchy horizon they cross, extrapolating to a step function drop at the Cauchy horizon. This behavior implies a diverging tidal force is experienced before any singularity (if present) is encountered. Similar conclusions were reached for the eternal 4D Kerr and Reissner-Nordstrom cases with a perturbative analysis ~\cite{marolf2012} and a few fully nonlinear case studies with numerics~\cite{eilon2016,Eilon:2016bcl,chesler2018,Chesler:2020lme}.

Statements about what the 3D AdS collapse case might say about the astrophysically relevant 4D black hole interior would be pure speculation. However, for some properties we already know there are qualitative differences between the two. For example, the work of Dafermos and Luk~\cite{dafermos2017} implies that in a black hole whose exterior approaches Kerr, the branch of the Cauchy horizon emanating from $i^+$ is weakly singular for all subextremal spins, unlike our high ($a\gtrsim 0.87$) spin cases. The work of Van de Moortel~\cite{VandeMoortel:2019ike,VandeMoortel:2020olr} shows a similar result for Reissner-Nordstrom, and moreover that the null Cauchy horizon (under reasonable assumptions) always ``breaks down'' to a central singularity, again in contrast to our high spin cases. It is unclear whether the latter difference is a consequence of charge versus angular momentum influencing the interior; a simple way to gain more insight would be to look at the interiors of black holes formed from charged circularly symmetric collapse in 3D AdS. Of course, the ultimate goal would be to study both 3D and 4D collapse with angular momentum and charge without any symmetry restrictions, though that would pose significant challenges for either analytic or numerical studies. The surprisingly rich set of outcomes found in the $AdS_3$ case, however---which is expected to be much simpler than the higher dimensional cases, where true dynamical gravitational degrees of freedom come into play---suggests that taking on the challenge would be well worth the effort.

\begin{acknowledgments}
We thank Mihalis Dafermos, David Garfinkle, Amos Ori, Jorge Santos and Maxime Van de Moortel for stimulating discussions related to this work.
This material is based upon work supported by the National Science Foundation (NSF) Graduate Research Fellowship Program under Grant No. DGE-1656466. Any opinions, findings, and conclusions or recommendations expressed in this material are those of the authors and do not necessarily reflect the views of the National Science Foundation. F.P. acknowledges support from NSF Grant No. PHY-1912171, the Simons Foundation, and the Canadian Institute For Advanced Research (CIFAR).
\end{acknowledgments}

\appendix

\section{Equations of Motion} \label{sec:EOM}
In this appendix we explicitly write down the particular variables we use, equations we solve, and the corresponding boundary and regularity conditions, closely following~\cite{jalmuzna2017}.

We use the metric ansatz (\ref{eq:jg_metric}), and scalar field ansatz (\ref{eq:jg_matter_ansatz}) with corresponding stress-energy tensor (\ref{eq:stress_energy}).
The Einstein field equations (\ref{efe}) can be decomposed into the Hamiltonian constraint
\begin{equation} \label{eq:Hamiltonian_constraint}
\begin{aligned}
&B'' + B' \Big(B' - A' + \frac{1 + \cos^2(r/\ell)}{\ell \sin(r/\ell) \cos(r/\ell)}\Big) \\
\ &- \frac{A'}{\ell \sin(r/\ell) \cos(r/\ell)} \\
&- \dot{A} \dot{B} + C_3 + \frac{1}{4} C_4 \gamma^2 + 4 \pi S_{B'} = 0,
\end{aligned}
\end{equation}
the radial component of the momentum constraint
\begin{equation} \label{eq:Momentum_constraint}
\begin{aligned}
&\dot{B}' + \dot{B} \Big(B'- A' + \frac{\cos(r/\ell)}{\ell \sin(r/\ell)}\Big) - \\
\ & \dot{A} \Big(B' + \frac{1}{\ell \sin(r/\ell) \cos(r/\ell)}\Big) + 4 \pi S_{\dot{B}} = 0,
\end{aligned}
\end{equation}
the angular component of the momentum constraint
\begin{equation} \label{eq:J_constraint}
J' + 8 \pi \bar{r} S_{\gamma'} = 0,
\end{equation}
and three independent components of the evolution equations
\begin{equation} \label{eq:B_evol_eqn}
\begin{aligned}
&-\ddot{B} + B'' + \frac{2}{r} B' + (B')^2 + B' \Big(\frac{2}{\ell \sin(r/\ell) \cos(r/\ell)} - \frac{2}{r}\Big) \\
&- \dot{B}^2 + 2 C_3 + \frac{1}{2} C_4 \gamma^2 + 4 \pi S_B = 0,
\end{aligned}
\end{equation}
\begin{equation} \label{eq:A_evol_eqn}
-\ddot{A} + A'' + C_3 - \frac{3}{4} C_4 \gamma^2 + 4 \pi S_A = 0,
\end{equation}
and
\begin{equation} \label{eq:J_evol_eqn}
\dot{J} + 8 \pi \bar{r} S_{\dot{\gamma}} = 0.
\end{equation}
In the above, the matter source terms are
\begin{equation} \label{eq:matter_source_terms}
\begin{aligned}
S_A, S_{B'} &= \frac{1}{2} \sin^{2}(r/\ell) [(X^2 + Y^2) \mp (V^2 + W^2)] \\
&+ \frac{1}{2 \ell^2} (\cos^2(r/\ell) \mp e^{2A - 2B}) (\phi^2 + \psi^2) \\
&+ \frac{\cos(r/\ell) \sin(r/\ell)}{\ell} (X \phi + Y \psi) \\
S_{B} &= \frac{1}{\ell^2} e^{2A - 2B} (\phi^2 + \psi^2) \\
S_{\dot{B}} &= \sin^{2}(r/\ell) (VX + WY) + \\
\ & \frac{\cos(r/\ell) \sin(r/\ell)}{\ell} (V \phi + W \psi) \\
S_{\dot{\gamma}} &= \sin^{2}(r/\ell) (Y \phi - X \psi) \\
S_{\gamma'} &= \sin^{2}(r/\ell) (W \phi - V \psi), 
\end{aligned}
\end{equation}
and we have introduced the following auxiliary variables:
$\gamma \equiv \beta'$, $X \equiv \phi'$, $Y \equiv \psi'$, $V \equiv \dot{\phi} +  \beta \psi$, $W \equiv \dot{\psi} - \beta \phi$, $C_1 \equiv 2/(\ell \tan(r/\ell)) + 1/(\ell \sin(r/\ell) \cos(r/\ell)) - 3/r + B'$,
$C_2 \equiv B'/(\ell \tan(r/\ell)) + (\cos^2(r/\ell) - e^{2 A - 2 B})/(\ell^2 \sin^2(r/\ell))$, $C_3 = (1 - e^{2 A})/(\ell^2 \cos^2(r/\ell))$, $C_4 = \ell^2 \sin^2(r/\ell) e^{2B - 2A}$, and $J \equiv \rb^3 \gamma/f$. 

In terms of the above first order variables, the Klein-Gordon equation for the two independent components of the complex scalar field take the form
\begin{equation} 
\begin{aligned} \label{eq:KG_eqn}
&-\dot{X} + V' - (\beta Y + \gamma \psi) = 0 \\
&-\dot{Y} + W' + (\beta X + \gamma \phi) = 0 \\
&-\dot{V} + X' + \frac{3}{r} X + C_1 X - \beta W - \dot{B} V + C_2 \phi = 0\\
&- \dot{W} + Y' + \frac{3}{r} Y + C_1 Y + \beta V - \dot{B} W + C_2 \psi = 0.
\end{aligned}
\end{equation}

We impose the following regularity conditions at the origin $r=0$
\begin{equation}
\begin{aligned}
A'(t,0) &= B'(t,0) = \beta'(t,0) = 0 \\
A(t,0) &= B(t,0) \\
\phi'(t,0) &= \psi'(t,0) = 0,
\end{aligned}
\end{equation}
and the following regularity/outer boundary conditions at $r = \ell \pi / 2$
\begin{equation}
\begin{aligned}
A(t, \ell\pi/2) &= B'(t, \ell\pi/2) = \beta(t, \ell\pi/2) = 0\\
\phi(t, \ell\pi/2) &= \psi(t, \ell\pi/2) = 0.
\end{aligned}
\end{equation}

\section{Convergence Tests} \label{sec:convergence_tests}

\begin{figure}[]
	\centering
	\includegraphics[width=0.5\textwidth,trim=0.3in 0.0in 0.0in 0.0in]{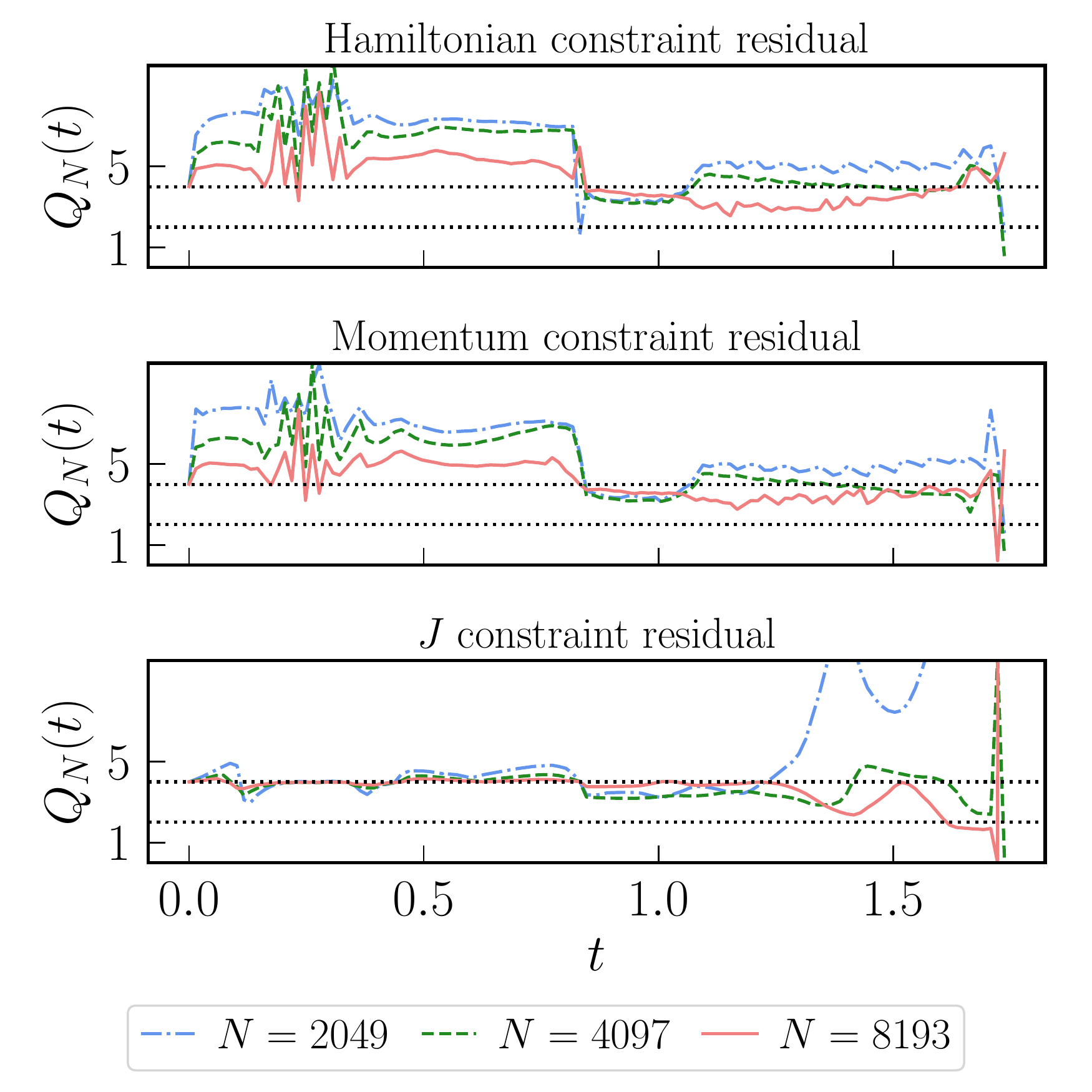}
	\caption{Plots of the rate of convergence to zero $Q_N(t)$ (\ref{eq:Q_res}) for the residuals of the three constraint equations (\ref{eq:Hamiltonian_constraint}-\ref{eq:J_constraint}), here for the case with $a = 0.77$.  Each should be converging to zero with $Q_N(t) = 4$ (in the continuum limit). We see a trend with increasing resolution to this expected behavior until very near the end, at which time most of the grid is excised and some field gradients have become quite large. In extracting properties from the numerical solutions for the results presented here, we do not use data from this ``noisy'' region; in particular all the extrapolation of quantities to the presumed spacelike singularity (when present), Cauchy horizon, and $i^+$ are performed with data in the region where we have good convergence. Dotted black lines are shown at $Q_N(t) = 2, 4$ to help guide the eye.}
	\label{fig:Q_res_plot}
\end{figure}

We have performed many tests to check the correctness of our code, including conservation of the asymptotic mass and angular momentum, and that the scheme is converging at the expected order. The latter rate of convergence should be second order throughout the evolution, though we sometimes see slightly better than second order for lower resolutions, presumably due to our use of fourth order spatial differences. For brevity, here we only show---in Fig.~\ref{fig:Q_res_plot}---one set of convergence tests from a representative case, namely convergence of the three constraint equations to zero for the $a=0.77$ case. As mentioned in the main text, we use a free evolution scheme, where the constraints are only solved at the initial time, and this is therefore a rather nontrivial test that we are solving the correct system of equations. Specifically, what is plotted in the figure are a set of ratios of the $L_2$ norms versus time of the residuals of each constraint, taken between pairs of successively higher resolution runs:
\begin{equation} \label{eq:Q_res}
Q_N(t) = \frac{||L^{2h} u^{2h}||}{||L^{h} u^{h}||},
\end{equation}
where $L^h u^h$ denotes a residual operator $L$ acting on the discrete solution 
$u$ at resolution (grid spacing) $h$ (and analogously for the half resolution case at $2h$), and $N$ is the number of points in the higher resolution run, related to $h$ by $h = (r_{max} - r_{min})/(N - 1) = (\ell \pi)/(2(N-1))$. In the continuum limit, $Q_N(t)$ should asymptote to $2^m$, where $m$ is the order of convergence.

\bibliography{references}

\end{document}